\documentclass[twocolumn, appendixfloats, revtex4, numberedappendix, twocolappendix, iop]{openjournal}

\usepackage{CJK}
\usepackage{enumitem}
\usepackage{url}
\usepackage{amsmath}
\usepackage[most,breakable]{tcolorbox}

\newcommand{\rev}[1]{#1}
\newcommand{\revtwo}[1]{#1}

\shorttitle{Egent: Autonomous EW Measurement}
\shortauthors{Ting et al.}

\begin{document}
\begin{CJK*}{UTF8}{gbsn}

\title{Egent: An Autonomous Agent for Equivalent Width Measurement \vspace{-1.5cm}}

\author{Yuan-Sen Ting (丁源森)}
\affiliation{Department of Astronomy, The Ohio State University, Columbus, OH 43210, USA}
\affiliation{Center for Cosmology and AstroParticle Physics (CCAPP), The Ohio State University, Columbus, OH 43210, USA}
\affiliation{Max-Planck-Institut f\"ur Astronomie, K\"onigstuhl 17, D-69117 Heidelberg, Germany \vspace{-0.3cm}}

\author{Serat Mahmud Saad}
\affiliation{Department of Astronomy, The Ohio State University, Columbus, OH 43210, USA \vspace{-0.3cm}}

\author{Fan Liu (刘凡)}
\affiliation{National Astronomical Observatories, Chinese Academy of Sciences, Beijing 100012, China \vspace{-0.3cm}}

\author{Yuting Shen}
\affiliation{School of Computer Science, Georgia Institute of Technology, Atlanta, GA 30332, USA}

\begin{abstract}
We present Egent, an autonomous agent that combines classical multi-Voigt profile fitting with large language model (LLM) visual inspection and iterative refinement. The fitting engine is built from scratch with minimal dependencies, creating an ecosystem where the LLM can reason about fits through function calls---adjusting wavelength windows, adding blend components, modifying continuum treatment, and flagging problematic cases. Egent operates directly on raw flux spectra without requiring pre-normalized continua. We validate against manual measurements from human experts using 18,615 lines from the C3PO program across 84 Magellan/MIKE spectra at SNR$\sim$50--250. \rev{The raw agreement between Egent and expert measurements is MAD~=~5--7~m\AA{}, without any post-hoc per-spectrum correction. Per-spectrum slopes of $\sim$0.85--1.19 around unity reflect differences in global continuum methodology rather than fitting failures.} The LLM's primary role is quality control: it confirms good fits ($\sim$60--65\% of lines are LLM-refined and accepted), flags problematic cases ($\sim$10--20\%), and occasionally rescues edge cases where tool use improves fits. Agreement between GPT-5 and GPT-5-mini confirms reproducibility, with GPT-5-mini enabling low-cost analysis at $\sim$200 lines per US dollar. Every fit stores complete Voigt parameters, continuum coefficients, and LLM reasoning chains, enabling exact reconstruction without re-running. Egent compresses what traditionally requires months of expert effort into days of automated analysis, enabling survey-scale EW measurement. We provide open-source code at \url{https://github.com/tingyuansen/Egent}, including a web interface for drag-and-drop analysis and a local LLM backend for fully offline operation on consumer hardware.
\end{abstract}

\section{Introduction}
\label{sec:intro}

Large spectroscopic surveys such as APOGEE \citep{Majewski2017}, GALAH \citep{Buder2025}, Gaia-ESO \citep{Randich2022}, and the ongoing SDSS-V \citep{Kollmeier2017}, 4MOST \citep{deJong2012}, and WEAVE \citep{Dalton2014} have propelled astronomy into the era of big spectroscopic data. The need for automated analysis has driven the adoption of full-spectrum fitting methods for stellar parameter and abundance determination \citep[e.g.,][]{Ness2015, Holtzman2015, Perez2016, Ting2019, Xiang2022}. Historically, however, spectra were analyzed through equivalent width (EW) measurements \citep[e.g.,][]{Korista1997, Milone2013, Montes2018}: individual spectral lines have characteristic shapes, and the integrated area under a line---the equivalent width---directly encodes the number of absorbing atoms in the stellar atmosphere, providing elemental abundances.

EW measurement persists for small-scale, high-precision studies. For high-resolution differential analysis, line-by-line EW measurement remains the standard approach, because individual spectral lines carry different systematic errors, including uncertain oscillator strengths \citep{Scott2015, Scott2015b}, unmodeled blending, and departures from local thermodynamic equilibrium \citep{Asplund2000, Asplund2009, Bergemann2019, Lind2024}. Full-spectrum fitting averages over these systematics, while differential EW analysis between similar stars cancels them. This line-by-line approach remains the consensus method for achieving precision abundance measurements \citep{Bensby2004, Bedell2014, Nissen2016}. Solar twin studies \citep{Ramirez2009}, planet engulfment signatures \citep{Galarza2021, Liu2024, Spina2024}, and precise elemental ratios all rely on it.

Existing tools range from fully interactive programs like IRAF's \texttt{splot}, where each line requires manual continuum placement and blend decisions, to automated codes like ARES \citep{Sousa2007}, DAOSPEC \citep{Stetson2008}, \rev{and TAME \citep{Kang2012}}, and integrated frameworks like iSpec \citep{BlancoCuaresma2014} and SMHR\footnote{https://github.com/andycasey/smhr} that combine EW measurement with abundance analysis. \rev{Beyond stellar spectroscopy, automated spectral decomposition has been tackled by tools such as GaussPy \citep{Lindner2015} and GaussPy+ \citep{Riener2019}, which use machine-learning-assisted Gaussian decomposition for ISM spectral line surveys (21~cm absorption and CO/H\,\textsc{i} emission, respectively); these share the goal of replacing manual inspection with algorithmic decisions, though they target different data regimes than stellar absorption-line work.} 

These codes can process line lists in batch mode, but they still require parameter calibration, typically expect pre-reduced spectra, and rely on visual examination of fits to identify problematic cases. \rev{The contribution of Egent is not multi-component fitting or heuristic quality control---both are well established---but the integration of an LLM visual inspector that reasons about diagnostic plots through function-calling tools, providing a quality-control loop that adapts to each fit without requiring hand-tuned rules.} For a typical differential abundance study of $\sim$100 stars with $\sim$200 lines each, an expert may spend months on measurements and quality checks.

\begin{figure*}[htbp!]
    \centering
    \includegraphics[width=2\columnwidth]{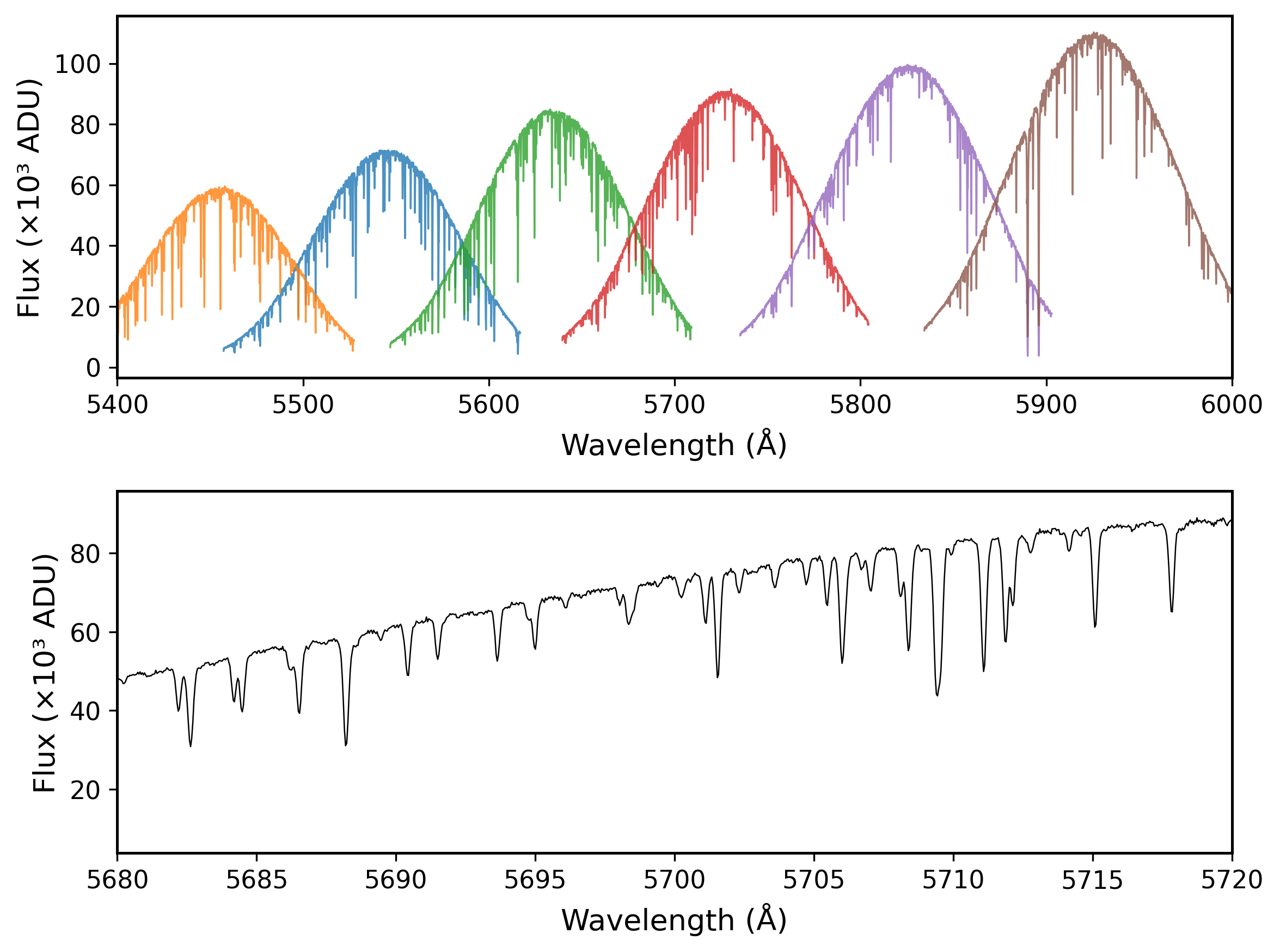}
    \caption{Raw Magellan/MIKE \'echelle spectrum illustrating challenges for automated EW measurement. \textit{Top panel:} Multiple \'echelle orders (different colors) showing the characteristic blaze function---the instrumental response that modulates the observed flux with a curved envelope peaking near each order's center. Absorption lines appear as narrow dips superimposed on this curved background. \textit{Bottom panel:} Single order zoom (5680--5720~\AA) showing raw flux without continuum normalization. The underlying continuum slopes upward following the blaze function, demonstrating why local continuum fitting is required. Visible challenges include blended lines (multiple absorptions within $<$1~\AA), weak lines near the noise level, and the varying continuum slope that traditional methods must remove before fitting.}
    \label{fig:echelle}
\end{figure*}

While in principle EW measurement involves fitting Voigt profiles to many lines simultaneously, full automation remains difficult---analogous to source extraction in imaging, where even with known profiles, edge cases abound. Simple algorithms fail on unrecognized blends, curved continuum from nearby strong lines, and crowded spectral regions \citep{Ramirez2009, Scott2015b, Liu2021}. Rule-based systems require extensive tuning for each spectrograph and stellar type, yet still miss unusual cases. Difficult lines have historically required human eyes and judgment.

A further complication is the instrumental response. Raw spectra do not show a flat continuum from the stellar photosphere; instead, the observed flux is modulated by the spectrograph's blaze function---a wavelength-dependent efficiency that varies smoothly across each \'echelle order \citep[][see Figure~\ref{fig:echelle}a]{Skoda2008}. This has led to extensive effort to ``continuum normalize'' spectra before fitting \citep{Ivanov2004, BlancoCuaresma2014, Perez2016}, introducing additional subjective choices that propagate into systematic errors.

Recent advances in large language models (LLMs) suggest a path forward. Modern LLMs exhibit robust reasoning capabilities \citep{Wei2022, achiam2023gpt}, and when equipped with function-calling tools, they become agents capable of taking actions \citep{Yao2022}, not merely generating text like a chatbot, but executing code, inspecting results, and iterating. Such agents have demonstrated autonomous scientific research capabilities in chemistry \citep{Boiko2023, Bran2023, Ramos2024}, materials science \citep{Szymanski2023}, and astronomy \citep{Sun2024,Sun2025}. For tasks like EW measurement, where the rules are difficult to encode exhaustively but natural to express in language (this blend looks unresolved, ``the continuum is tilted''), an LLM agent can fill the gap that rule-based automation cannot.

This paper demonstrates that an LLM-based agent can provide the judgment traditionally requiring human experts. We build an EW measurement pipeline from scratch, creating an ecosystem of tools that the LLM can invoke: fitting functions, diagnostic plots, continuum adjustments, and quality flags. By fitting locally within each wavelength window, Egent avoids the need for global continuum normalization---the LLM handles continuum placement as part of its iterative refinement. All decisions are logged, providing complete provenance that manual measurements lack. Egent occupies a middle ground: classical Voigt fitting (fast, interpretable) with LLM-guided decisions for difficult cases, uniquely providing complete provenance for every measurement.

\section{Method: The Egent Pipeline}
\label{sec:method}

\subsection{Design Philosophy}

For an agentic pipeline, we need tools that the agent can invoke---mimicking how a human expert would interact with traditional software like IRAF: zooming in on a spectral region, choosing line centers, deciding continuum placement, adding blend components. Our initial approach was to wrap IRAF itself (via PyRAF) and let the LLM agent drive it, but this proved difficult given the many interconnected components and legacy dependencies.

Instead, we recognized that for EW fitting specifically, the core task reduces to multi-component Voigt profile fitting---something straightforward to implement from scratch. Egent is built in pure Python using only standard scientific libraries (\texttt{numpy}, \texttt{scipy.optimize}, \texttt{scipy.special} for the Faddeeva function). We deliberately avoid legacy packages and complex dependencies; everything is self-contained and inspectable, apart from the LLM API calls---the interface through which our code sends prompts to and receives responses from external language model services.\footnote{For an introduction to LLM APIs and tool use, see \url{https://tingyuansen.github.io/coding\_essential\_for\_astronomers/}.}

This design serves multiple purposes. First, it ensures reproducibility across computing environments with minimal installation requirements---particularly important for the offline version we present later. Second, version-controlled dependencies and a simple codebase make agentic tool use more robust; complex environments with many packages introduce failure modes that degrade the agent's reliability. Third, the transparent fitting algorithm allows users to understand and modify the analysis.

The core principle of Egent is that physics defines the fitting; the LLM operates through tools. The LLM does not perform calculations or invent new methods. It inspects diagnostic plots, identifies problems that a human expert would notice, and suggests parameter changes using the same tools a human would use. This division of labor leverages the LLM's strength (visual pattern recognition and reasoning) while keeping the physics grounded in well-understood spectral line models.

We primarily use GPT-5-mini for production runs, with GPT-5 serving as a cross-check for validation. Both models support function calling and vision capabilities required for our pipeline. GPT-5-mini processes approximately 200 lines per dollar (averaging $\sim$5,000 input tokens and $\sim$1,000 output tokens per line across multiple iterations), while GPT-5 is roughly five times more expensive ($\sim$40 lines per dollar). The consistency between models (demonstrated in Section~\ref{sec:results}) validates that our pipeline does not rely on model-specific behaviors, and the cost efficiency of GPT-5-mini makes large-scale surveys economically feasible.

\begin{figure*}
    \centering
    \includegraphics[width=2\columnwidth]{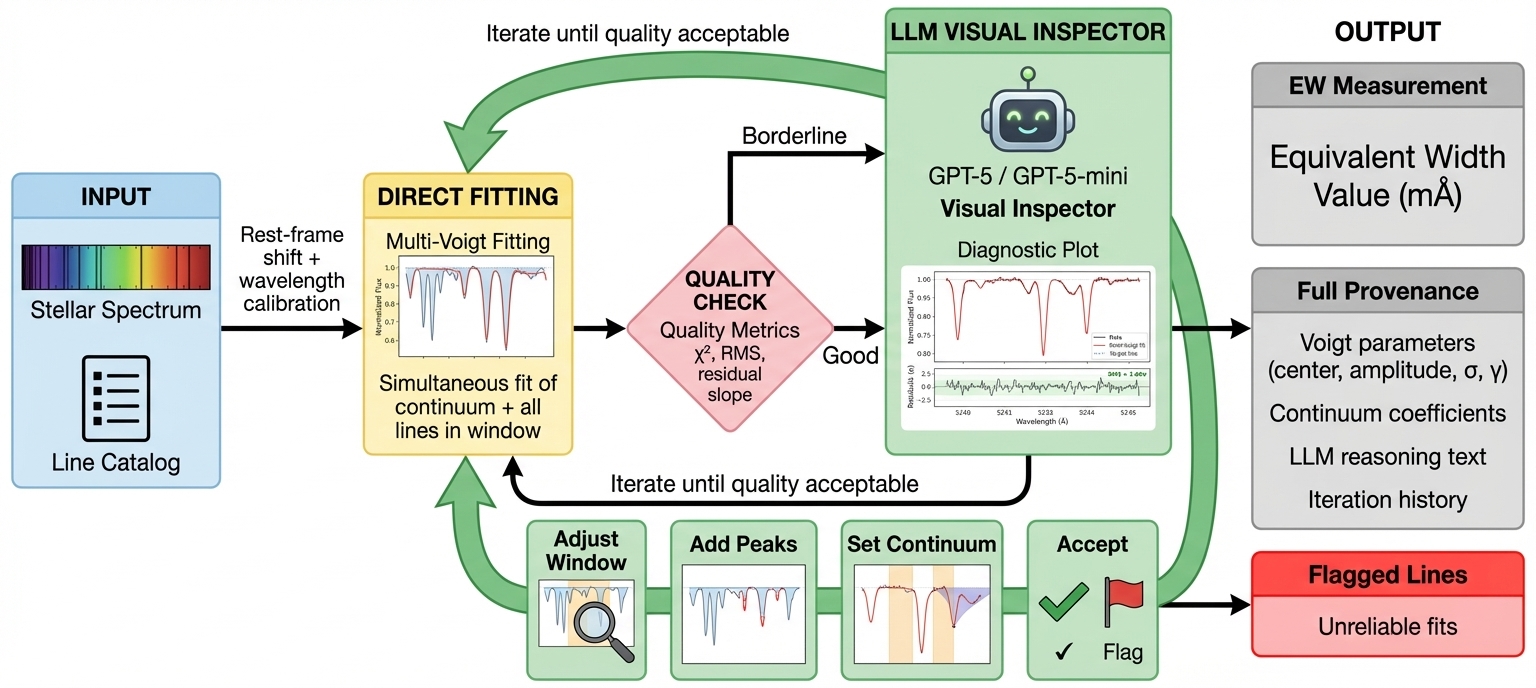}
    \caption{Schematic overview of the Egent pipeline. Input spectra (shifted to stellar rest frame with barycentric and empirical wavelength corrections) and line catalogs enter the direct multi-Voigt fitting stage. Quality metrics determine whether the fit is acceptable or requires LLM inspection. The LLM visual inspector examines diagnostic plots and can adjust the extraction window (Fig.~\ref{fig:llm_improvement}), add peaks for blends (Fig.~\ref{fig:peak_identification}), propose continuum regions based on visual reasoning (Fig.~\ref{fig:continuum_regions}), or flag unreliable fits. The iteration loop continues until the LLM accepts the measurement. All outputs include complete provenance: Voigt parameters, continuum coefficients, and LLM reasoning chains.}
    \label{fig:schema}
    \end{figure*}

\subsection{Pipeline Overview}

Figure~\ref{fig:schema} illustrates the complete Egent workflow. \rev{Egent expects input spectra already in the stellar rest frame} (e.g., after applying barycentric and radial-velocity corrections). Egent requires a user-provided line list specifying target wavelengths to measure; for each target line in this list, the pipeline proceeds as follows:

\begin{enumerate}[leftmargin=*]
\item Extract a local spectral region around the target wavelength (default $\pm$3~\AA).
\item Identify all catalog lines within this window that might contribute to the observed absorption.
\item Fit a multi-Voigt model plus polynomial continuum simultaneously to the extracted region (Section~\ref{sec:multivoigt}).
\item Compute quality metrics (Section~\ref{sec:quality}) and either accept the fit directly or trigger LLM visual inspection (Section~\ref{sec:llm_inspection}).
\end{enumerate}

\rev{The default window of $\pm$3~\AA\ is chosen to be large relative to the line broadening kernel. At $R = 50{,}000$ (Magellan/MIKE), the instrumental FWHM is $\lambda/R \approx 0.11$~\AA\ near 5500~\AA, so the window spans $\sim$55 resolution elements---sufficient to capture full Voigt wings and neighboring blends. For slowly rotating solar-type stars ($v\sin i \lesssim 5$~km\,s$^{-1}$), rotational broadening adds $\lesssim 0.1$~\AA, which remains well within the window. The LLM can adapt the window during iterative refinement: narrowing it (e.g., to $\pm$2~\AA) for isolated lines where edge features contaminate the continuum fit, or widening it (to $\pm$5~\AA) when strong lines are truncated.}

A key insight is that simple scalar metrics like $\chi^2$ are insufficient for quality assessment. A fit may have excellent global $\chi^2$ while completely missing a blend at the target wavelength, or conversely, have elevated $\chi^2$ due to edge effects that do not affect the target measurement. What matters is whether the model correctly captures the absorption \textit{at and around the target line}---a judgment that requires inspecting the spatial pattern of residuals, not just their aggregate statistics. This is precisely the kind of visual pattern recognition that LLMs with vision capabilities now excel at.

The recent development of multimodal LLMs with strong performance on non-natural images---including scientific plots---makes this approach timely. Earlier vision models struggled with plots, charts, and diagrams; current models have improved ability to interpret axis labels, identify systematic residual patterns, and reason about what they imply for fit quality.

The following subsections detail each component: the equivalent width definition (Section~\ref{sec:ew_def}), the Voigt profile model (Section~\ref{sec:voigt}), multi-component fitting (Section~\ref{sec:multivoigt}), the fitting procedure (Section~\ref{sec:fitting}), quality metrics (Section~\ref{sec:quality}), LLM visual inspection (Section~\ref{sec:llm_inspection}), and stored output (Section~\ref{sec:output}).

\subsection{Equivalent Width Definition}
\label{sec:ew_def}

The equivalent width is a classical concept in stellar spectroscopy, measuring integrated line absorption relative to the continuum:
\begin{equation}
W_\lambda = \int \frac{F_c(\lambda) - F(\lambda)}{F_c(\lambda)} \, d\lambda
\end{equation}
where $F_c(\lambda)$ is the continuum flux and $F(\lambda)$ is the observed flux. For a fitted line profile $\phi(\lambda)$ normalized such that the continuum equals unity:
\begin{equation}
W_\lambda = \int [1 - \phi(\lambda)] \, d\lambda
\end{equation}

Note that this definition does not inherently require pre-normalized spectra. Continuum normalization---with its myriad tools, subjective choices, and operator-dependent parameters---is often what makes EW measurements difficult to reproduce between researchers. Egent sidesteps this entirely. At high spectral resolution, the local continuum spanning a few angstroms can be well approximated by a low-order polynomial, often just linear. By fitting the continuum simultaneously with the line profile within each local window, we avoid the systematic errors and reproducibility issues introduced by global continuum normalization.

\subsection{The Voigt Profile}
\label{sec:voigt}

Spectral lines are broadened by multiple physical mechanisms, making the Voigt profile---the convolution of a Gaussian (thermal broadening) and a Lorentzian (pressure broadening)---the appropriate model for stellar absorption lines:
\begin{equation}
V(\lambda; \lambda_0, A, \sigma, \gamma) = A \cdot \frac{\text{Re}[w(z)]}{\sigma\sqrt{2\pi}}
\end{equation}
where $w(z)$ is the Faddeeva function and $z = \frac{(\lambda - \lambda_0) + i\gamma}{\sigma\sqrt{2}}$. The Gaussian width $\sigma$ captures thermal broadening (from the Maxwell-Boltzmann velocity distribution of absorbing atoms) and instrumental broadening (the spectrograph's line spread function). The Lorentzian width $\gamma$ captures natural broadening (from the finite lifetime of the excited state) and pressure broadening (collisions with other atoms). Rotational broadening can also be approximated as Gaussian for the moderate resolutions and slowly-rotating stars considered here; for rapid rotators at very high resolution, explicit rotational kernels would be a straightforward addition. We evaluate the Voigt profile using \texttt{scipy.special.voigt\_profile}.

\begin{figure*}
    \includegraphics[width=\textwidth]{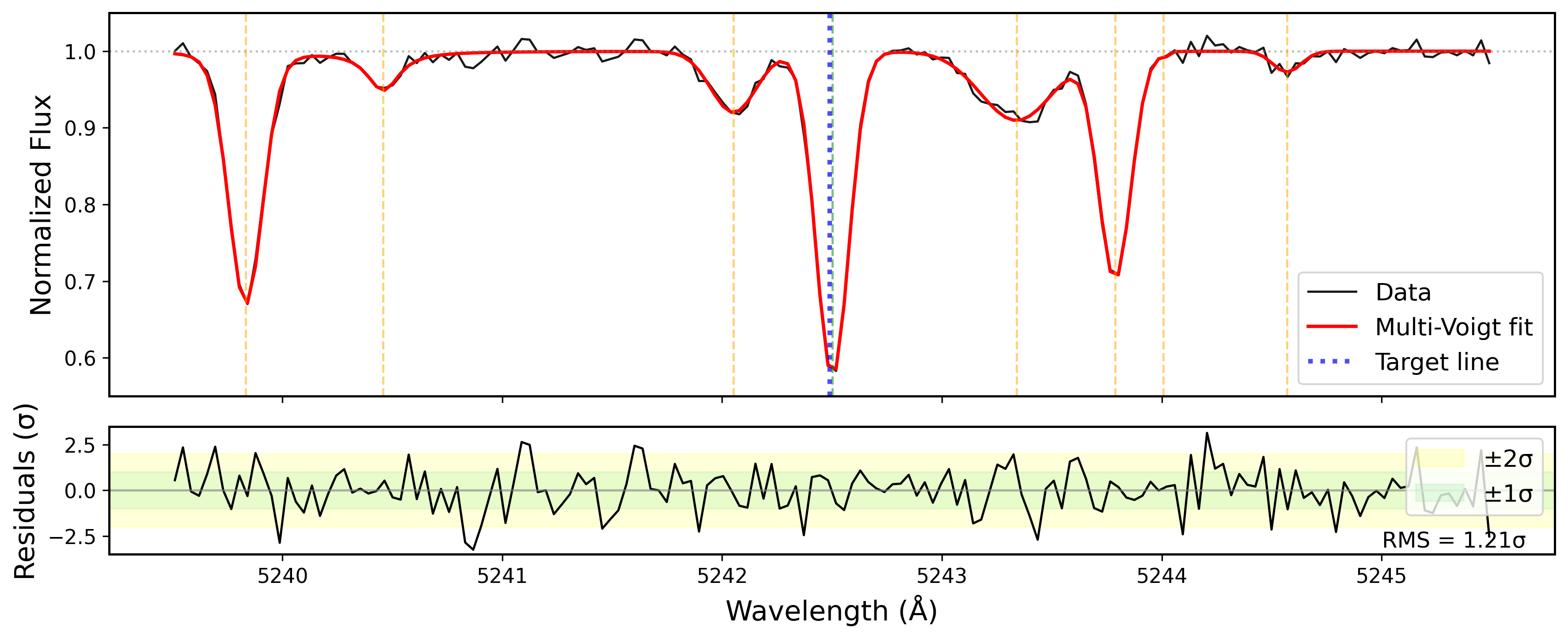}
    \caption{Multi-Voigt fitting process demonstrated on Fe~I 5242.49~\AA\ (Gaia ID 55780840513067392, SNR $\sim$ 160). \textit{Top panel:} The normalized spectrum (black) is fit simultaneously with 8 Voigt components (orange and green dashed lines indicate individual component centers; the target component is shown in green). The combined multi-Voigt model (red) reproduces the complex blend structure. Continuum normalization is performed iteratively alongside the Voigt fitting, allowing the model to adapt to local continuum curvature. \textit{Bottom panel:} Normalized residuals (data minus model, divided by flux uncertainties) demonstrate fit quality. The shaded regions indicate $\pm 1\sigma$ (green) and $\pm 2\sigma$ (yellow); the RMS of 1.21$\sigma$ confirms a statistically acceptable fit. When the linear fit to residuals shows a slope $>$0.3~$\sigma$/\AA, a red dashed trend line is overlaid to guide the LLM's attention to continuum issues. This diagnostic panel format is also used for LLM visual inspection, where the agent examines residual patterns to detect missed blends (W-shaped residuals), continuum issues (systematic slopes indicated by the trend line), or unreliable fits (excessive scatter near the target line). \rev{Residuals are shown in $\sigma$ units throughout; to convert to absolute normalized-flux units, multiply by the per-pixel uncertainty $\sigma_{\rm pixel} \approx 1/\text{SNR}$ (e.g., for this spectrum at SNR~$\sim$~160, 1$\sigma$ corresponds to $\sim$0.006 in normalized flux).}}
    \label{fig:multivoigt}
\end{figure*}

\subsection{Multi-Component Fitting}
\label{sec:multivoigt}

A limitation of traditional manual EW measurement is fitting one line at a time. An experienced spectroscopist may identify obvious blends and fit them together, but this is labor-intensive---and even careful human analysis rarely accounts for all lines within the fitting window, each of which affects the derived continuum level. Ignoring these neighbors leads to systematic errors, particularly in crowded spectral regions.

Egent addresses this by fitting all lines in the extraction window simultaneously. The number of components $N$ is initially determined from the input line catalog: all catalog entries within the extraction window are included. During LLM refinement, the agent can propose additional peaks if it identifies unmodeled absorption features in the residuals (Fig.~\ref{fig:peak_identification}). The model is:
\begin{equation}
F_{\rm model}(\lambda) = C(\lambda) \cdot \left[1 - \sum_{i=1}^{N} V_i(\lambda; \lambda_{0,i}, A_i, \sigma_i, \gamma_i)\right]
\end{equation}
where $C(\lambda)$ is the continuum and the sum runs over all $N$ lines in the window. Each Voigt profile $V_i$ has four parameters: center wavelength $\lambda_{0,i}$, amplitude $A_i$, Gaussian width $\sigma_i$, and Lorentzian width $\gamma_i$.

The continuum $C(\lambda)$ is parameterized as a polynomial centered on the mean wavelength of the extraction window. By default, we use a linear continuum:
\begin{equation}
C(\lambda) = c_0 + c_1 \cdot (\lambda - \bar{\lambda})
\end{equation}
For regions with curved continuum (e.g., near the wing of a strong line), the LLM can request a higher-order polynomial:
\begin{equation}
C(\lambda) = \sum_{k=0}^{n} c_k \cdot (\lambda - \bar{\lambda})^k
\end{equation}
with $n = 2$ (quadratic) or $n = 3$ (cubic). Higher orders are available but rarely needed and can lead to overfitting. The LLM makes this determination by inspecting residual patterns: a systematic curve in the residuals suggests the need for higher-order continuum.

The continuum level is determined from pixels identified as line-free. Initially, this is done automatically by iterative sigma-clipping: pixels more than $2.5\sigma$ below the running median are masked as absorption, and the continuum is fit to the remaining points. If the LLM identifies problems---for example, in a crowded region where no pixels are truly line-free---it can visually identify wavelength intervals that appear relatively absorption-free and propose these as continuum anchors (Fig.~\ref{fig:continuum_regions}).

The total number of free parameters is $(n+1)$ for the continuum plus $4N$ for the Voigt profiles. For a typical extraction window containing 3--5 lines with linear continuum, this corresponds to 14--22 parameters fitted to several hundred data points.

\subsection{Fitting Procedure}
\label{sec:fitting}

The fitting procedure begins with initial parameter guesses: line centers from the input catalog, amplitudes estimated from the local flux minimum near each catalog position, and Gaussian and Lorentzian widths initialized to typical values appropriate for the spectrograph resolution ($\sigma = 0.04$~\AA, $\gamma = 0.015$~\AA).

Optimization is performed using the Levenberg-Marquardt algorithm via \texttt{scipy.optimize.curve\_fit}. This provides both the best-fit parameters and their covariance matrix, from which we estimate parameter uncertainties. \rev{After fitting, components whose centers converge to within 0.2~\AA\ ($\approx$1.8 resolution elements at $R = 50{,}000$ near 5500~\AA) are merged, as they cannot be meaningfully resolved and their parameters become degenerate. The covariance matrix from \texttt{curve\_fit} is stored in the output JSON for every accepted fit, enabling users to assess parameter correlations and fit stability post hoc.} If the fit fails to converge, we attempt restarts with perturbed initial conditions before flagging the line.

After fitting, we validate that no two Voigt profile centers have converged to within 0.2~\AA\ of each other---a situation indicating that two components are modeling the same absorption feature (overfitting). If such pairs are found, we remove the component farther from the target wavelength and refit with fewer peaks. This ensures each absorption feature is modeled by a single Voigt profile, preventing double-counting of equivalent width.

The equivalent width is then computed by numerical integration of the target line's Voigt profile. When multiple lines are fitted simultaneously, we report the EW for the component whose center is closest to the catalog wavelength.

\subsection{Quality Metrics}
\label{sec:quality}

We compute several diagnostics to assess fit quality and decide whether LLM review is needed.

The reduced chi-squared quantifies overall goodness of fit:
\begin{equation}
\chi^2_\nu = \frac{1}{N_{\rm dof}} \sum_i \frac{(F_i - F_{\rm model,i})^2}{\sigma_i^2}
\end{equation}
where $N_{\rm dof}$ is the number of data points minus the number of free parameters. Values near unity indicate a good fit; larger values suggest the model is inadequate.

The residual RMS, normalized by flux uncertainties, provides a more intuitive measure:
\begin{equation}
\text{RMS} = \sqrt{\frac{1}{N} \sum_i \left(\frac{F_i - F_{\rm model,i}}{\sigma_i}\right)^2}
\end{equation}
where $\sigma_i$ is the flux uncertainty at each pixel. For a statistically good fit, this should be approximately unity; values above 1.5--2 indicate systematic deviations beyond what noise alone would produce.

We also compute a linear fit to the normalized residuals, yielding a residual slope in units of $\sigma$/\AA. Slopes exceeding $\sim$0.5~$\sigma$/\AA\ indicate a systematic continuum problem that requires attention. When the slope exceeds 0.3~$\sigma$/\AA, we overlay the linear trend on the diagnostic plot (Figure~\ref{fig:multivoigt}b), giving the LLM a direct visual cue for continuum issues. We additionally check for correlated residuals (which might indicate a missed blend or continuum curvature) and the offset between the fitted and catalog wavelengths (which might indicate a misidentified line).

\rev{The pipeline classifies fits into three tiers. Fits with RMS~$\lesssim 2\sigma$ and no quality issues are classified as \textit{good} and accepted directly. Fits with moderate diagnostics (e.g., RMS~$\sim$~2--3$\sigma$, or 5--9 lines in the window) are classified as \textit{acceptable} and also accepted without LLM intervention. LLM visual inspection is triggered when any of the following conditions are met: overall RMS~$> 3\sigma$, $\chi^2_\nu > 5$, $|\text{residual slope}| > 0.5~\sigma$/\AA, $\geq$5 consecutive residual points exceeding $2\sigma$, fitted center offset~$> 0.3$~\AA\ from catalog, central RMS~$> 2.5\sigma$ (within $\pm$1.5~\AA\ of the target), or $\geq$10 lines in the extraction window. These thresholds can be relaxed (or tightened) to send more (fewer) lines to the LLM at higher (lower) cost. A complete listing of all configuration parameters is provided in Table~\ref{tab:config}.}

\subsection{LLM Visual Inspection}
\label{sec:llm_inspection}

When quality metrics trigger inspection, the LLM receives a diagnostic plot showing the data, fitted model, individual Voigt components, and residuals (Figure~\ref{fig:multivoigt}b). The plot is annotated with the target wavelength and quality metrics. The LLM examines the plot as an expert spectroscopist would, looking for specific problems:

\begin{itemize}[leftmargin=*]
\item W-shaped residuals centered on a wavelength offset from the target indicate a missed blend. The LLM can propose adding a Voigt component at the approximate wavelength.
\item Tilted or systematically biased residuals across the window indicate a continuum problem. The LLM can adjust the extraction window or switch to a polynomial continuum.
\item Large scatter in the residuals near the target wavelength, or an offset fitted center, may indicate an unreliable measurement. The LLM can flag the line as unreliable.
\end{itemize}

\noindent The LLM communicates these actions through a set of function-calling tools (Appendix~\ref{app:schemas}) that mirror the operations a human spectroscopist would perform: adjusting the wavelength window, adding blend components, changing continuum models, or flagging problematic lines.

After proposing an action, the pipeline executes it, re-fits, and returns a new diagnostic plot. The LLM visually assesses whether the fit has improved and either accepts the measurement or proposes further refinement. This iteration continues until the LLM accepts the fit or a maximum iteration count (default 10) is reached.

The LLM operates as a blind analysis: it has no access to the catalog EW values. All decisions are based solely on the fit diagnostics and visual appearance of the plot. This prevents the LLM from adjusting parameters to match expected values.

Below we show several representative examples of LLM reasoning on typical EW measurement cases (see Box 1 for verbatim outputs from GPT-5-mini):

\begin{tcolorbox}[colback=black!3, colframe=black!50, title=Case 1: Blend Detection (Ca~I 6696.02~\AA), fonttitle=\bfseries, breakable]
\small
``The residuals show a pronounced W-shaped pattern centered at the target (a positive excursion just blueward of 6696.02~\AA\ and negative residuals at/just redward), indicating a missed blend very close to the target. The red fit looks symmetric while the data shows asymmetry (suggesting two close components). I will add an extra Voigt component to deblend the pair at the blue-side shoulder near 6695.9~\AA.''\\[0.3em]
\textit{After adding peak:} ``The residuals no longer show a W-shape at the target and the central RMS is 1.64$\sigma$, which is within the acceptable range.''
\end{tcolorbox}
\begin{tcolorbox}[colback=black!3, colframe=black!35, title=Case 2: Window + Polynomial Continuum (Fe~I 6261.10~\AA), fonttitle=\bfseries, breakable]
\small
``The fit shows very high normalized RMS (7.07) and many correlated residuals; the residuals show a strong systematic tilt across the target region indicating the continuum fit is wrong. I will try a polynomial continuum to better model curvature.''\\[0.3em]
\textit{After polynomial + narrower window:} ``After adding nearby blends, switching to a 2nd-order polynomial continuum, and narrowing the window to 2.0~\AA, the residuals near the target are flat with no W-shaped patterns and the central RMS is 1.23$\sigma$. I will accept this measurement.''
\end{tcolorbox}
\begin{tcolorbox}[colback=black!3, colframe=black!65, title=Case 3: Manual Continuum Regions --- Flagged (Si~I 5665.55~\AA), fonttitle=\bfseries, breakable]
\small
``The region is very crowded with multiple absorptions. Automatic continuum fitting showed a large slope and correlated residuals. I will force the continuum using manually selected line-free regions that appear flat in the plot: $\sim$5663.25--5663.60~\AA\ (left side), $\sim$5664.70--5665.05~\AA\ (just blue of target), and $\sim$5668.20--5668.50~\AA\ (right side).''\\[0.3em]
\textit{Final assessment:} ``Despite adding peaks and manual continuum regions, the fitter still reports very high RMS. The surrounding blends and systematic residuals make the EW unreliable. I will flag as severe\_blend.''
\end{tcolorbox}

The approach balances thoroughness against computational cost: direct acceptance for clearly good fits (saving LLM costs), and iterative refinement with visual inspection for borderline cases where human-like judgment is needed.

\begin{figure*}
    \centering
    \includegraphics[width=0.85\textwidth]{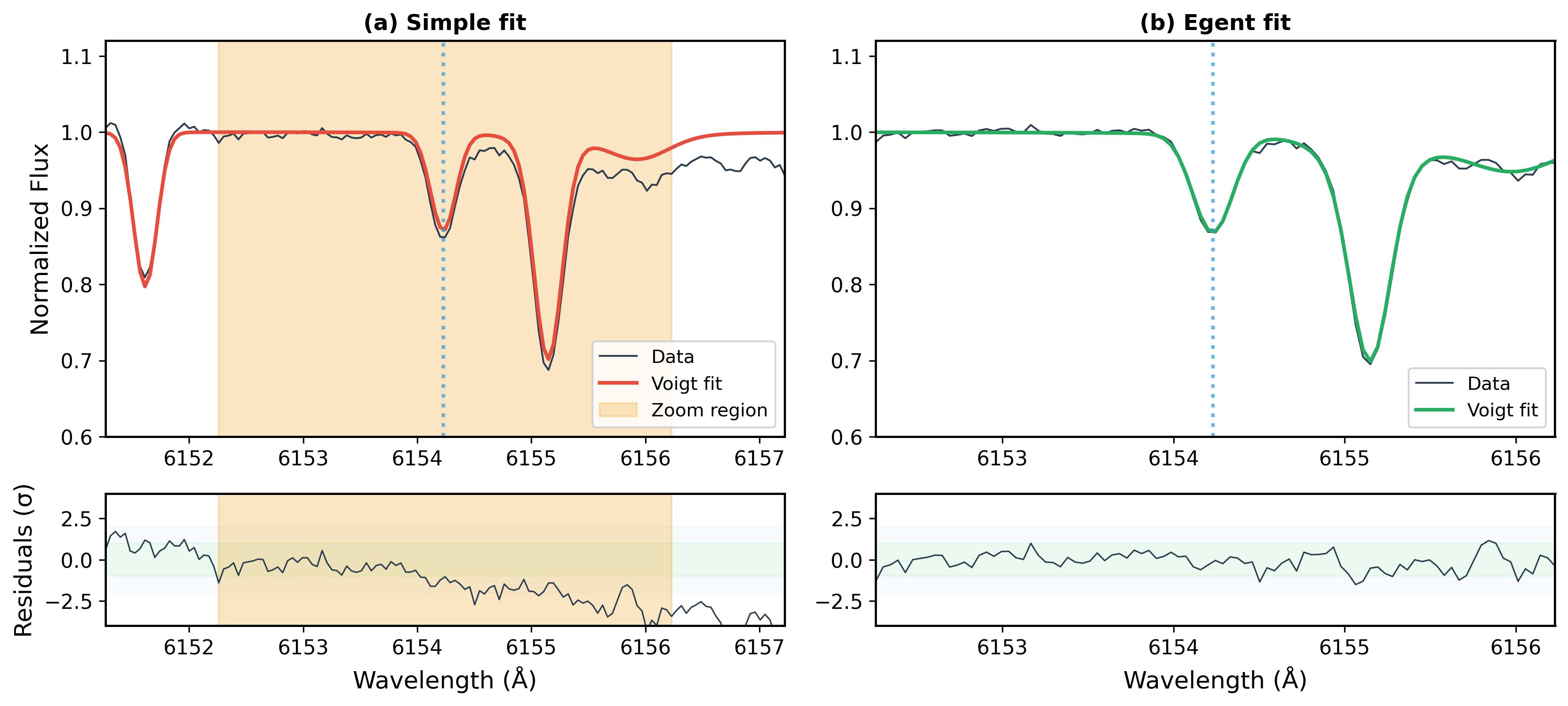}
    \caption{Example of LLM-driven window and continuum adjustment for Na~I 6154.23~\AA. \textbf{(a) Simple fit:} The initial 6~\AA\ window includes edge features (visible at 6156--6157~\AA) that introduce systematic bias in the residuals. The yellow shaded region indicates where the LLM determined a narrower window would improve the fit. \textbf{(b) Egent fit:} After the agent narrowed the window to 3.9~\AA\ and switched to a quadratic continuum, the fit focuses on the region containing the target lines. The residuals improve: systematic deviations outside the $\pm 1\sigma$ band (green shading) largely disappear. This case illustrates the agent's ability to reason about fitting strategy: by excluding edge features irrelevant to the target line and adjusting the continuum model, the measurement becomes more robust. The EW error decreased from 8.9~m\AA\ to 5.8~m\AA\ (35\% improvement relative to the C3PO catalog value of 36.8~m\AA).}
    \label{fig:llm_improvement}
\end{figure*}

\subsection{Stored Output}
\label{sec:output}

Every measurement is stored as a JSON record with complete provenance, enabling exact reconstruction of the fit without re-running the LLM. Each record includes:

\begin{itemize}[leftmargin=*]
\item \textbf{Measurements:} Final EW and uncertainty, direct (pre-LLM) EW for comparison, quality classification, and difference from catalog if available
\item \textbf{Voigt parameters:} For every fitted line in the window---center wavelength, amplitude, $\sigma$, $\gamma$, FWHM, and individual EW---enabling exact model reconstruction
\item \textbf{Continuum:} Method used (iterative linear, polynomial, or manual regions), polynomial order, and fitted coefficients
\item \textbf{Region:} \'Echelle order, wavelength range, extraction window size
\item \textbf{Diagnostics:} Reduced $\chi^2$, normalized RMS, quality issues and warnings, number of fitted lines
\item \textbf{LLM provenance:} Boolean indicating whether LLM was triggered, complete conversation transcript with reasoning text before each action, iteration history showing how EW evolved across refinements
\item \textbf{Metadata:} Processing time, path to diagnostic plot, method used (direct/LLM), flag status and reason if applicable
\end{itemize}

This level of detail serves multiple purposes: it allows verification that specific measurements are reliable, enables post-hoc replotting without re-fitting, provides training data for future fine-tuned models, and establishes a standard for reproducibility that manual measurements cannot match. The explicit reasoning text (Box~1) documents \textit{why} the LLM made each decision, not just what it did.

\begin{table*}
\caption{\rev{Default configuration parameters for the Egent pipeline. Left columns list extraction and fitting parameters that control the Voigt profile model and continuum estimation. Right columns list quality-control thresholds that determine whether a fit is accepted directly or sent for LLM visual inspection, along with pipeline settings. Fits not meeting any LLM trigger condition are accepted directly. All values can be modified by users; see the code repository for details.}}
\label{tab:config}
\rev{
\setlength{\tabcolsep}{0.6em}
\begin{tabular*}{\textwidth}{@{\extracolsep{\fill}} lr|lr}
\hline\hline
\multicolumn{2}{c|}{\textbf{Extraction \& Fitting}} & \multicolumn{2}{c}{\textbf{Quality \& LLM Triggers}} \\
\hline
Window half-width & 3.0~\AA & LLM trigger: high RMS & Overall RMS $> 3.0\sigma$ \\
Continuum method & Iterative linear & LLM trigger: high $\chi^2$ & $\chi^2_\nu > 5$ \\
Polynomial order & 1 (LLM can raise) & LLM trigger: central RMS & Central RMS $> 2.5\sigma$ \\
Sigma-clip threshold & 2.5$\sigma$ & LLM trigger: residual slope & $\lvert\mathrm{slope}\rvert > 0.5~\sigma$/\AA \\
Top percentile (init) & 85\% & LLM trigger: correlated resid. & $\geq$5 consecutive $>2\sigma$ \\
Min.\ peak height & 0.02 & LLM trigger: centre offset & $> 0.3$~\AA\ from catalog \\
Min.\ peak prominence & 0.015 & LLM trigger: crowded region & $\geq$10 lines in window \\
$\sigma$ init / bounds & 0.04 / [0.003, 0.25]~\AA & Trend line overlay & slope $> 0.3~\sigma$/\AA \\
$\gamma$ init / bounds & 0.015 / [0.001, 0.08]~\AA & Max LLM iterations & 10 per line \\
Centre bounds & $\pm$0.5~\AA\ from catalog & Max LLM time & 300~s per line \\
Centre merge threshold & 0.2~\AA & Flag retry attempts & 3 \\
\cline{3-4}
Max fit evaluations & 20{,}000 & \multicolumn{2}{c}{\textbf{Pipeline}} \\
\cline{3-4}
 & & Default model (API) & GPT-5-mini \\
 & & Default model (local) & Qwen3-VL-8B-4bit \\
 & & Parallel workers & 10 (API) / 1 (local) \\
\hline\hline
\end{tabular*}
}
\end{table*}

\section{Validation Data}
\label{sec:data}

Validating autonomous EW measurement requires a homogeneous dataset with expert measurements performed by a single analyst using consistent methodology. Synthetic spectra, while offering known ground truth, lack realistic complications such as instrumental blaze functions, telluric contamination, and the subtle blending patterns present in real stellar atmospheres. Conversely, measurements compiled from multiple sources suffer from analyst-to-analyst and methodology-to-methodology variations that confound meaningful comparison. The ideal benchmark combines real observational data with careful, single-analyst measurements---a combination that is rare given the labor-intensive nature of manual EW work.

We validate Egent against the C3PO (Complete Census of Co-moving Pairs Of stars) program \citep{Yong2023,Liu2024,SunQ2025,SunQ2025b,Yu2025}, a systematic study of 125 co-moving stellar pairs designed to probe chemical homogeneity in comoving systems and detect signatures of planet engulfment. The program obtained high-resolution spectra using three instruments: Magellan/MIKE (78 pairs, $R \sim 50{,}000$), Keck/HIRES (23 pairs, $R \sim 72{,}000$), and VLT/UVES (25 pairs, $R \sim 110{,}000$). D.~Yong, with about 20 years of experience in precision abundance work, performed all EW measurements using the line-by-line differential technique that achieves relative abundance precision of $\sim$0.01~dex. Each spectrum includes measurements for approximately 200 carefully selected lines spanning Fe~I, Fe~II, and 20+ other species. The C3PO measurements have enabled detection of chemical abundance differences at the 0.01~dex level between co-moving pairs, demonstrating their quality for precision differential analysis.

We analyze Magellan/MIKE spectra from the C3PO sample. For GPT-5-mini, we processed 63 spectra (13,913 lines); for GPT-5, we processed 21 spectra (4,702 lines). These numbers provide sufficient statistics to characterize both random scatter and systematic offsets. The spectra span signal-to-noise ratios of 50--250 per pixel near 5500~\AA\ (median 174) and cover wavelengths from 3300--9400~\AA, providing a representative range of data quality encountered in precision abundance studies.

We emphasize that expert measurements are not absolute ground truth. The ``true'' EW depends on subjective choices about continuum placement and blend handling that different experts would make differently. However, expert measurements from a single experienced analyst provide the fairest benchmark for real stellar spectra, as they incorporate the same challenges---blending, continuum ambiguity, noise---that any automated system must handle. An expert's years of experience inform subtle judgments about which blends are problematic and which continuum regions are trustworthy, judgments that idealized synthetic tests cannot capture.\footnote{D.~Yong has since transitioned out of active research.}

\rev{Egent assumes input spectra are in the stellar rest frame; it does not perform its own wavelength calibration.} Spectra are shifted to the stellar rest frame using barycentric corrections computed from observation timestamps and observatory coordinates. We apply an additional empirical wavelength correction using $\sim$10 strong, isolated Fe~I lines per spectrum, fitting observed positions against laboratory wavelengths to correct residual offsets ($\sim$0.01--0.05~\AA) arising from spectrograph flexure and reduction pipeline limitations. \rev{For the benchmark sample, we selected the subset of spectra for which this two-step correction (barycentric + Fe~I empirical shift) places absorption lines at their laboratory wavelengths to within the fitting tolerance ($\lesssim$0.3~\AA, the centre-bound half-width in Table~\ref{tab:config}). This selection does not bias the EW comparison, as EW is an integrated quantity insensitive to small wavelength offsets provided the line falls within the extraction window.}

\begin{figure*}
    \centering
    \includegraphics[width=0.85\textwidth]{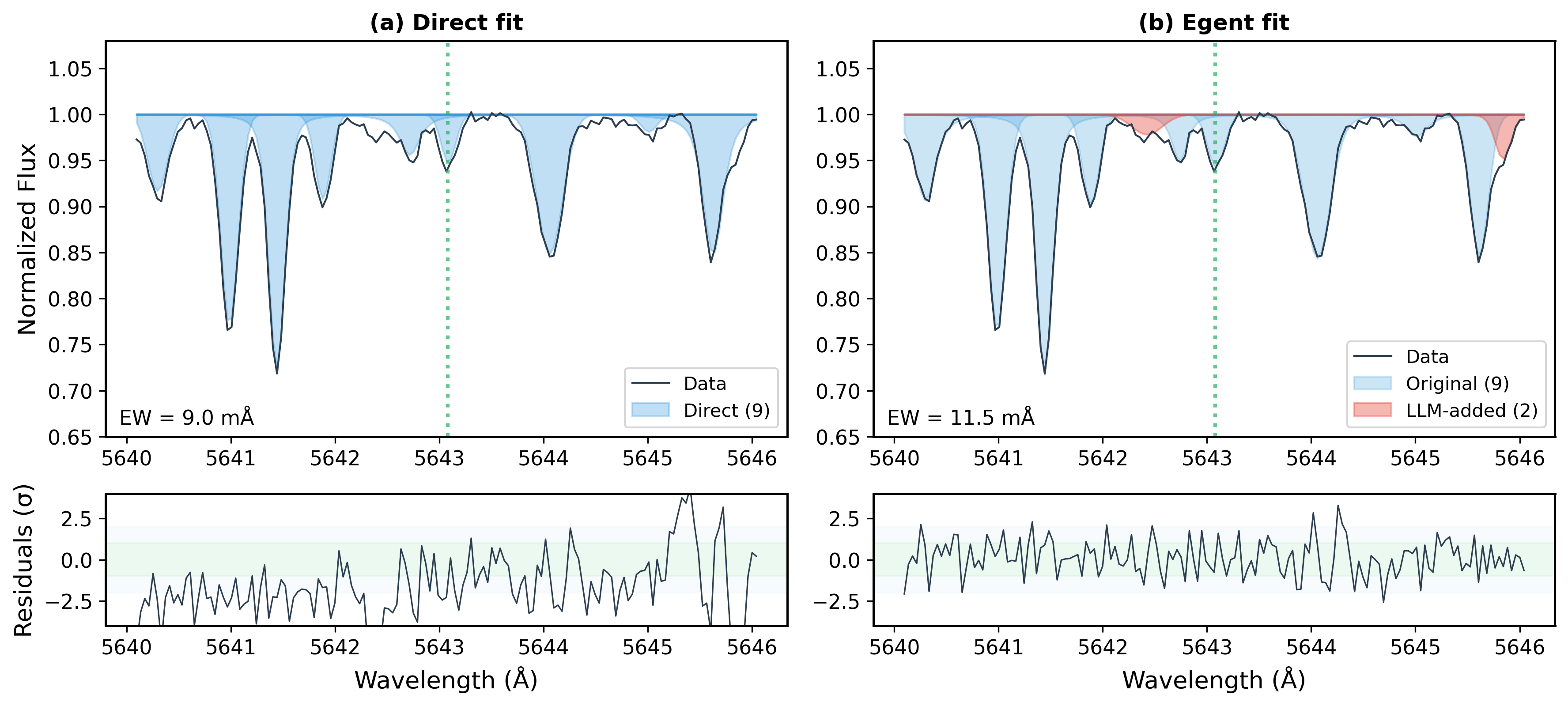}
    \caption{Example of LLM-driven peak identification for Ni~I 5643.08~\AA. \textbf{(a) Direct fit:} Nine Voigt components (blue shaded regions) were identified automatically from the line catalog. While these capture most absorption features, residuals at $\sim$5642.5~\AA\ and $\sim$5645.8~\AA\ show systematic deviations exceeding $2\sigma$, indicating missed features. \textbf{(b) Egent fit:} By visually inspecting the diagnostic plot, the LLM identified two small absorption features not in the catalog and explicitly specified their wavelengths (5642.45 and 5645.85~\AA) for inclusion in the multi-Voigt fit. The resulting 11-component model (9 original in blue, 2 LLM-identified in red) achieves flatter residuals. This demonstrates the LLM's ability to function as a visual inspector: it examines the residual panel for patterns indicating unmodeled absorption and proposes specific wavelengths for additional Voigt components. The EW improved from 9.0~m\AA\ to 11.5~m\AA, reducing the error relative to the catalog value (13.8~m\AA) by 52\%.}
    \label{fig:peak_identification}
    \end{figure*}
    
    \begin{figure*}
    \centering
    \includegraphics[width=1.7\columnwidth]{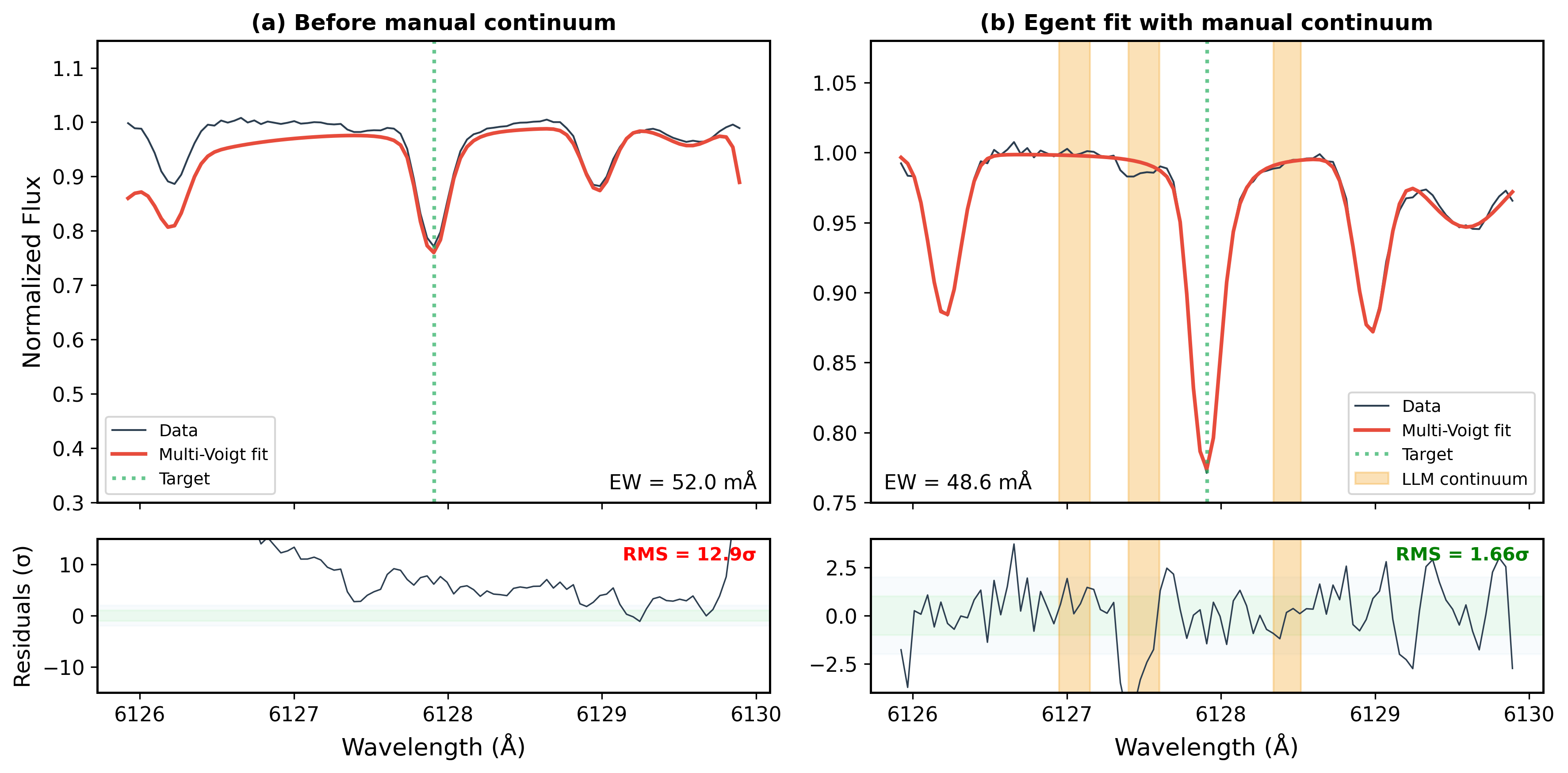}
    \caption{Before and after comparison of LLM-proposed continuum regions for Fe~I 6127.91~\AA. \textbf{(a) Before:} Automatic iterative continuum fitting struggles in this crowded region, resulting in a systematic offset between data and model (RMS~=~12.9$\sigma$). The normalized flux appears above unity on the blue side while absorption features are incompletely modeled. \textbf{(b) After:} Based on visual inspection, the LLM identified three wavelength intervals (orange shaded regions: 6126.95--6127.15~\AA, 6127.40--6127.60~\AA, and 6128.34--6128.52~\AA) as relatively line-free and proposed these as continuum anchors. With these constraints, the fitting procedure achieves improved residuals (RMS~=~1.66$\sigma$). This approach mimics how a human expert manually identifies clean continuum pixels when automatic methods fail in crowded spectral regions.}
    \label{fig:continuum_regions}
\end{figure*}

\section{Results}
\label{sec:results}

We present results in three parts. First, we examine the impact of LLM intervention: how often it triggers, what tools it uses, building intuition for how it improves fits. Second, we compare Egent measurements against the C3PO expert catalog, examining both aggregate statistics and individual case studies (Section~\ref{sec:comparison}). Third, we assess robustness across SNR using the natural variation in our sample (SNR$\sim$50--250), finding that performance remains stable with only modest degradation at the lowest SNR (Section~\ref{sec:snr_test}).

\begin{figure*}
    \centering
    \includegraphics[width=2\columnwidth]{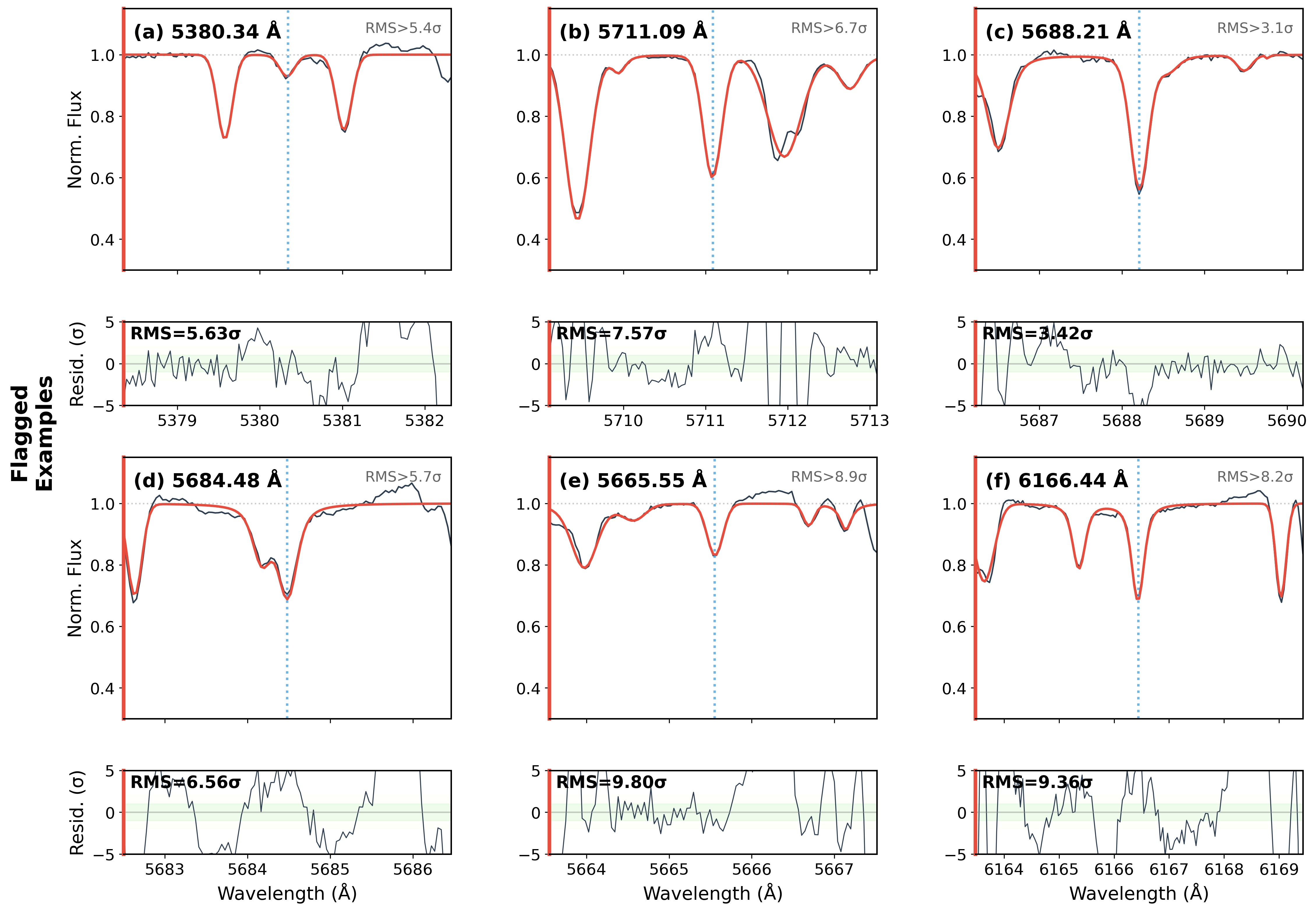}
    \caption{Representative examples of flagged lines. Lines are flagged by the LLM during iterative fitting when it determines the fit cannot be salvaged---elevated RMS values indicate continuum failures, severe blends, or model-data mismatch. The flagging mechanism identifies cases where automated fitting produces unreliable measurements, preventing these outliers from contaminating abundance analyses. All flagging uses AI visual inspection with no human intervention.}
    \label{fig:flagged_examples}
 \end{figure*}

\subsection{Impact of LLM Corrections}
\label{sec:llm_impact}

Table~\ref{tab:llm_stats} summarizes the pipeline behavior across GPT-5 and GPT-5-mini runs. Of the 4,702 lines processed with GPT-5 (21 spectra), 21.4\% were accepted on the direct fit without LLM intervention, while 59.4\% triggered LLM visual inspection and were subsequently accepted. For GPT-5-mini \rev{(63 spectra passing the $R^2 \geq 0.8$ correlation filter described below, totalling 13,913 lines)}, the pattern is similar: 26.3\% direct acceptance, 62.5\% LLM-triggered. The flagging rates show GPT-5 flagged 19.1\% of lines while GPT-5-mini flagged 11.0\%. The final acceptance rates are 80.9\% for GPT-5 and 88.8\% for GPT-5-mini.

\begin{table}
\centering
\caption{Pipeline statistics for GPT-5 and GPT-5-mini runs.}
\label{tab:llm_stats}
\begin{tabular}{lcc}
\hline
Metric & GPT-5 & GPT-5-mini \\
\hline
Spectra processed & 21 & 63 \\
Total lines & 4,702 & 13,913 \\
Direct fit accepted & 21.4\% & 26.3\% \\
LLM triggered \& accepted & 59.4\% & 62.5\% \\
Flagged & 19.1\% & 11.0\% \\
Final accepted & 80.9\% & 88.8\% \\
\hline
\end{tabular}
\end{table}

Both models follow a similar workflow: inspect the diagnostic plot, assess quality, and either accept or intervene. Table~\ref{tab:tool_usage} shows the usage frequency of each intervention tool. The key tools are \texttt{set\_continuum\_method} (switching to polynomial continuum), \texttt{extract\_region} (adjusting the wavelength window), \texttt{set\_continuum\_regions} (manually specifying line-free pixels), and \texttt{fit\_ew} with \texttt{additional\_peaks} (adding blend components). GPT-5-mini uses \texttt{set\_continuum\_regions} more frequently (7\% of LLM lines vs 1\% for GPT-5), while GPT-5 uses \texttt{set\_continuum\_method} and \texttt{extract\_region} more aggressively. These differences may reflect model-specific reasoning strategies, though both achieve comparable final accuracy.

\begin{table}[h]
\centering
\caption{Intervention tool usage (percentage of LLM-triggered lines). Percentages sum to $>$100\% because multiple tools are often used per line. The higher rates for GPT-5 partly reflect an earlier code version where the LLM explicitly called setup tools.}
\label{tab:tool_usage}
\begin{tabular}{lcc}
\hline
Tool & GPT-5 & GPT-5-mini \\
\hline
\texttt{set\_continuum\_method} & 49\% & 28\% \\
\texttt{extract\_region} & 84\% & 24\% \\
\texttt{fit\_ew} (re-fit) & 71\% & 32\% \\
\texttt{set\_continuum\_regions} & 1\% & 7\% \\
\hline
\end{tabular}
\end{table}

Figures~\ref{fig:llm_improvement}--\ref{fig:continuum_regions} illustrate the three main types of LLM intervention.

\textbf{Window and continuum adjustment} (Figure~\ref{fig:llm_improvement}): For Na~I 6154.23~\AA, the initial 6~\AA\ window included strong features at the edges (6156--6157~\AA) that pulled the linear continuum fit away from the true local continuum. The LLM recognized this pattern in the residuals---systematic deviations at the window edges---and responded by (1) narrowing the extraction window to 3.9~\AA\ to exclude the edge contamination, and (2) switching to a quadratic continuum to better model the local curvature. These two adjustments reduced the EW discrepancy from 8.9~m\AA\ to 5.8~m\AA\ (35\% improvement).

\begin{figure*}
    \includegraphics[width=2\columnwidth]{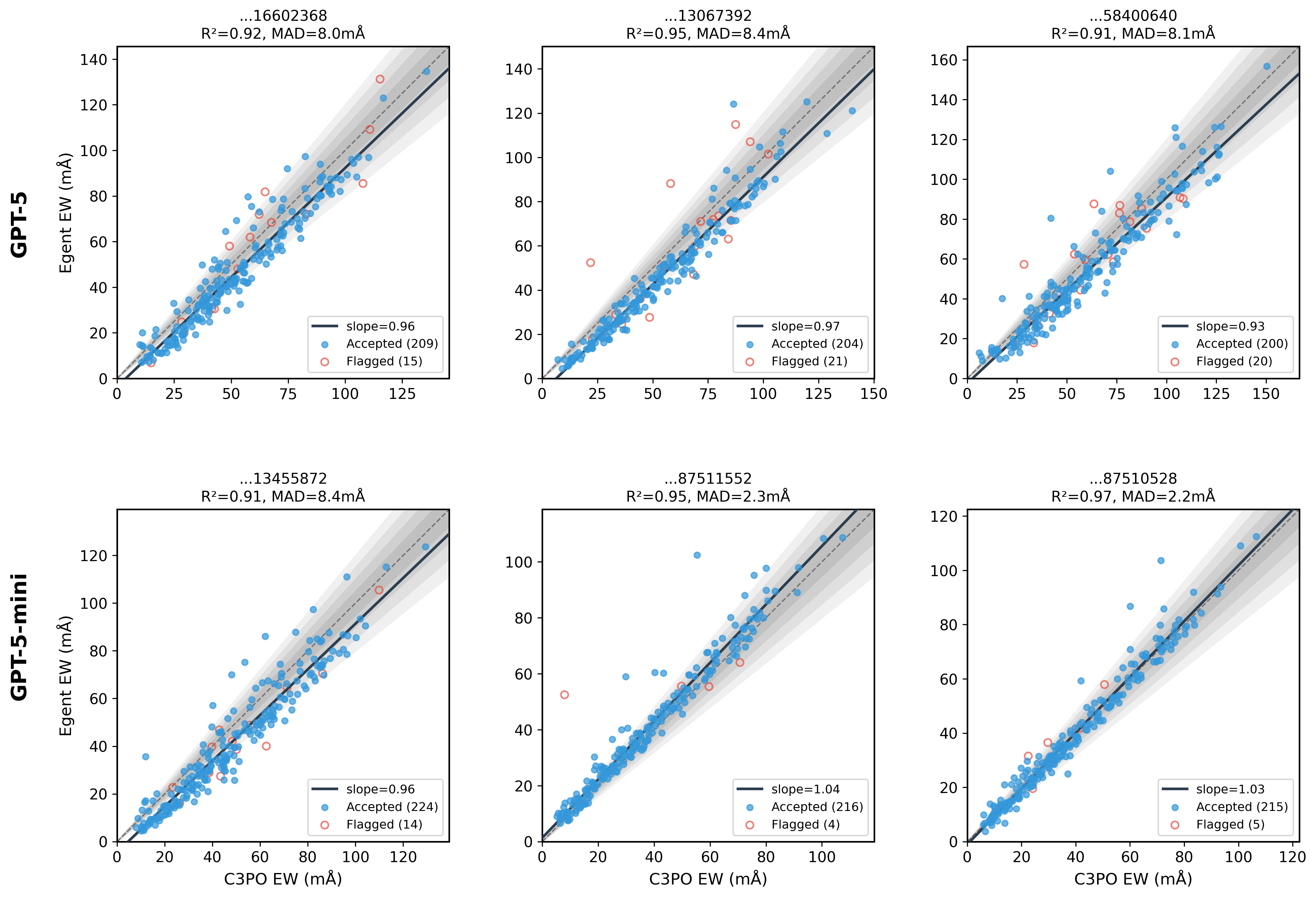}
    \caption{1-to-1 comparison of Egent vs C3PO EW measurements. Top row: GPT-5 results for three representative spectra. Bottom row: GPT-5-mini results. Blue filled circles: accepted measurements. Hollow red circles: lines flagged by the LLM. Dark line: best-fit linear regression on unflagged points only. Shaded bands: 5\%, 10\%, 15\%, 20\% deviation from 1:1. Flagged points cluster at larger deviations, validating the AI flagging mechanism. The per-spectrum slope variations ($\pm$5--10\%) arise from C3PO's global pre-normalization step---a subjective choice that introduces systematic offsets. Egent avoids this by fitting directly from raw flux with local continuum estimation, eliminating global continuum as a source of systematic error. Both models achieve $R^2 > 0.9$ and MAD~$\sim$~4--6~m\AA.}
    \label{fig:onetoone}
\end{figure*}

\textbf{Blend identification} (Figure~\ref{fig:peak_identification}): For Ni~I 5643.08~\AA, the direct fit included nine Voigt components from the catalog, but residuals at $\sim$5642.5~\AA\ and $\sim$5645.8~\AA\ showed systematic W-shaped patterns---the telltale signature of unmodeled absorption. The LLM identified these patterns visually and proposed two additional Voigt components at specific wavelengths (5642.45 and 5645.85~\AA). The 11-component fit achieved flatter residuals. This demonstrates the LLM's ability to detect blends.

\textbf{Manual continuum regions} (Figure~\ref{fig:continuum_regions}): For Fe~I 6127.91~\AA\ in a crowded spectral region, automatic iterative continuum fitting failed entirely (RMS = 12.9$\sigma$). No amount of sigma-clipping could identify true continuum pixels because absorption features dominated the entire window. The LLM recognized this failure mode and switched to manual continuum specification, visually identifying three narrow wavelength intervals (6126.95--6127.15~\AA, 6127.40--6127.60~\AA, and 6128.34--6128.52~\AA) that appeared relatively line-free. Using only these regions as continuum anchors, the fit improved to RMS = 1.66$\sigma$. This tool is invoked in 1--7\% of LLM-triggered lines (Table~\ref{tab:tool_usage}) but is essential for the most challenging spectral regions.

While Figures~\ref{fig:llm_improvement}--\ref{fig:continuum_regions} demonstrate improvements in specific edge cases, statistical analysis reveals the LLM's primary role is quality control. For GPT-5-mini, 80\% of LLM-triggered lines show final EW differing from the direct fit by $<$1~m\AA---the LLM examines the fit and confirms it is acceptable without modification. GPT-5 is more interventionist: only 59\% of its LLM-triggered lines have $<$1~m\AA\ changes, with 32\% showing $>$2~m\AA\ shifts (vs.\ 16\% for GPT-5-mini). This aligns with GPT-5's higher flagging rate and slightly better agreement with catalog values (Figure~\ref{fig:onetoone}). The LLM's value lies in (1) confirming that the majority of direct fits are reliable without human inspection, (2) flagging the 11--19\% of fits that cannot be salvaged, and (3) rescuing edge cases like those shown in Figures~\ref{fig:llm_improvement}--\ref{fig:continuum_regions} that would otherwise produce gross errors.

The LLM's value is most apparent in difficult cases. For the 1--7\% of lines requiring manual continuum specification (\texttt{set\_continuum\_regions}), automatic fitting fails entirely in crowded regions. For catastrophic fits with RMS $>$5$\sigma$, missed blends, or wrong continuum, LLM intervention can reduce errors from tens of m\AA\ to a few m\AA\ (see Figures~\ref{fig:llm_improvement}--\ref{fig:continuum_regions} for examples). These rescues represent only a few percent of all lines, but without them such cases would contribute catastrophic outliers to abundance analyses.

\rev{To quantify what the LLM adds beyond deterministic diagnostics, we compare the full pipeline against a direct-fit-only ablation. Without the LLM, applying the same RMS~$> 3\sigma$ flagging threshold to the direct fit accepts 10,134 lines at MAD~=~5.4~m\AA; with the LLM, 12,359 lines are accepted at MAD~=~5.8~m\AA. The 22\% gain in completeness comes from 2,424 lines that the LLM rescues: their initial direct-fit RMS exceeded 3$\sigma$, but the LLM improved the fit (via continuum adjustment, window narrowing, or blend addition) to bring residuals below threshold. These rescued lines have MAD~=~7.0~m\AA---worse than average but scientifically usable. Of the 9,602 LLM-triggered inspections, 71\% result in no EW change ($<$0.1~m\AA), confirming the direct fit without modification---the LLM's role in these cases is purely confirmatory. Routing only the top 5\% of hardest cases (ranked by direct-fit RMS) to the LLM already captures most of the MAD improvement, suggesting a tiered approach can achieve similar quality at $\sim$80\% cost savings. The LLM's contribution is thus primarily one of \textit{completeness}---rescuing marginal fits that deterministic rules would discard---and \textit{nuanced flagging} that accounts for spatial residual patterns beyond what scalar diagnostics capture. Since abundance determinations average over many lines per element, this 22\% gain in accepted lines translates directly into improved abundance precision.}
    
\begin{figure*}
    \includegraphics[width=2\columnwidth]{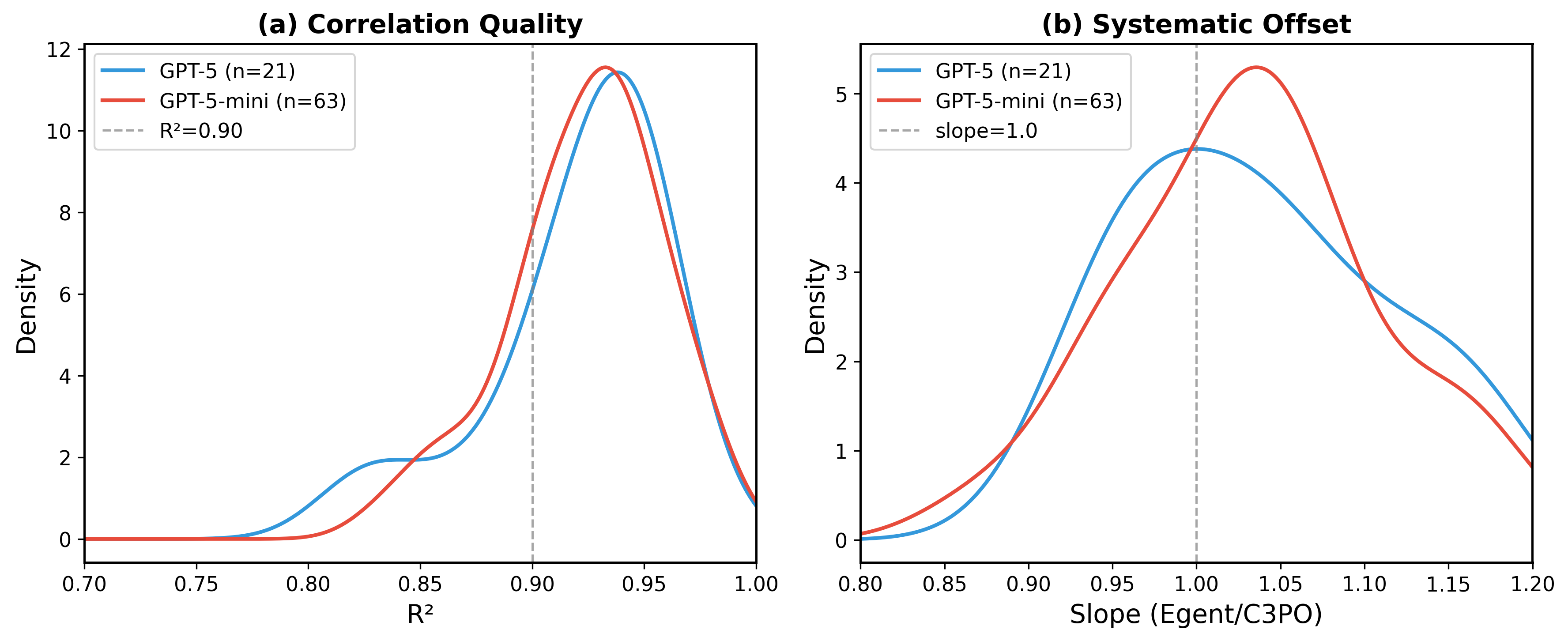}
    \caption{Distribution of correlation quality ($R^2$) and systematic offset (slope) across all spectra with C3PO catalog matches, comparing GPT-5 (21 spectra) and GPT-5-mini (63 spectra). \rev{A quality gate of $R^2 \geq 0.8$ is applied to the per-spectrum linear regression of Egent EW vs.\ C3PO EW (computed on unflagged lines with $\geq$20 matched lines); all spectra in the pre-selected sample pass this threshold (see text).} (a) Both models achieve median $R^2 \approx 0.93$, with all spectra above 0.82, indicating strong linear correlation between Egent and expert measurements. (b) Slopes are centered near unity (median $\approx$ 1.03) with $\pm$15\% spread, reflecting per-spectrum offsets from continuum methodology differences rather than fitting failures. The overlapping distributions demonstrate that GPT-5-mini achieves measurement quality comparable to GPT-5 at 5$\times$ lower cost.}
    \label{fig:r2_slope}
\end{figure*}

\begin{figure*}
    \centering
    \includegraphics[width=2\columnwidth]{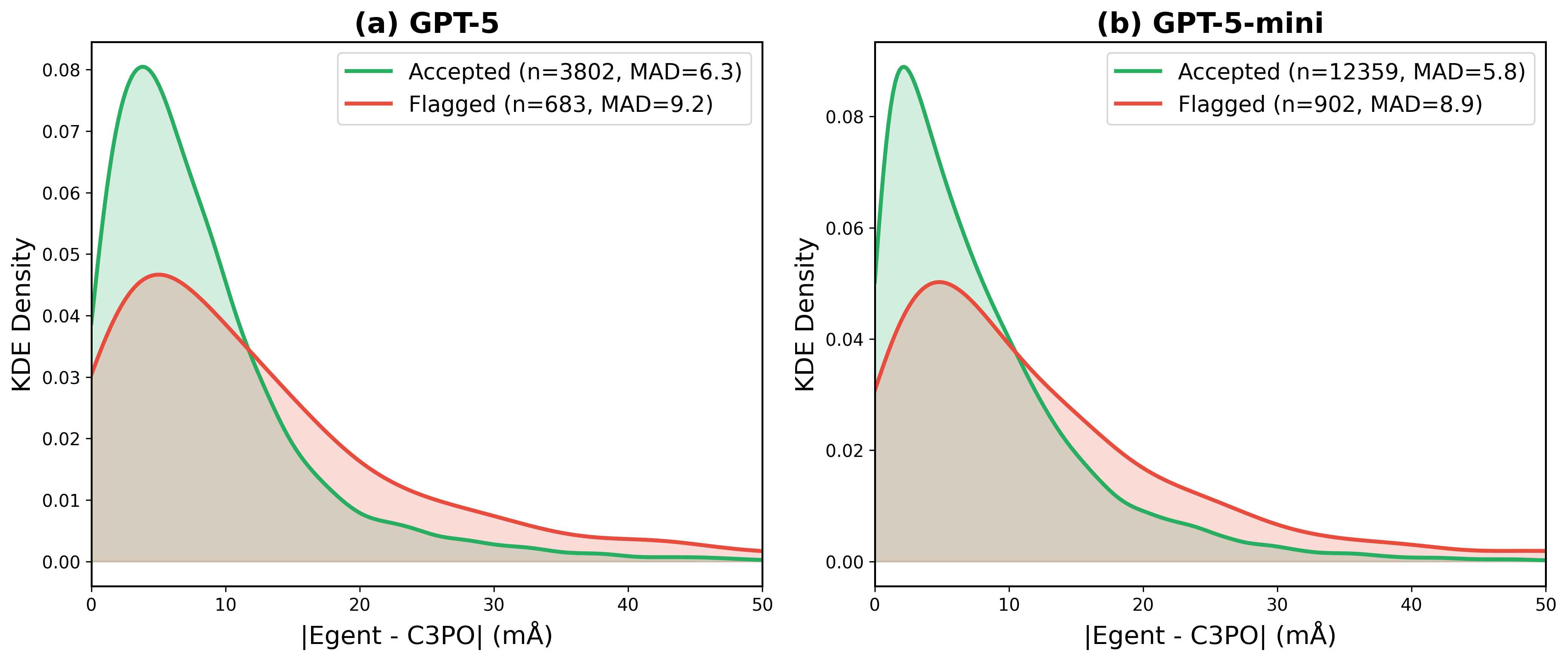}
    \caption{Distribution of absolute EW deviations $|$Egent $-$ C3PO$|$ separated by flagging status for (a) GPT-5 (21 spectra, 4,702 lines) and (b) GPT-5-mini (63 spectra, 13,913 lines). Green: accepted lines (MAD = 5--7~m\AA). Red: flagged lines (MAD = 10--20~m\AA). Both models show the same pattern: flagged lines have systematically larger deviations from catalog values, confirming that the AI flagging effectively identifies problematic fits.}
    \label{fig:deviation_by_flag}
\end{figure*}

\textbf{AI flagging} (Figure~\ref{fig:flagged_examples}): The pipeline employs fully automated quality control with no human intervention. During the iterative fitting process, the LLM can invoke the \texttt{flag\_line} tool when it determines that a fit cannot be salvaged after multiple improvement attempts. Figure~\ref{fig:flagged_examples} shows representative cases where elevated RMS values indicate catastrophic failures (wrong continuum, severe blends, or persistent model-data mismatch). The LLM is explicitly instructed to focus on the target region ($\pm$0.5~\AA\ of the target wavelength) and ignore edge effects that do not affect the measurement. This focused approach ensures thorough quality control without over-flagging lines with acceptable fits at the target.

\rev{Evaluated as a binary classifier using $|\Delta\text{EW}| > 10$~m\AA\ as the operational definition of a ``poor'' measurement, the flagging system achieves precision $\approx$~46\% and recall $\approx$~11\% for GPT-5-mini (23\% for GPT-5). The false-negative rate among accepted lines is $\sim$28\% at this threshold. These modest classifier metrics reflect the distinction between \textit{fit quality} (what flagging targets, based on residual patterns) and \textit{agreement with a reference catalog} (what $|\Delta\text{EW}|$ measures). A large $|\Delta\text{EW}|$ can arise from legitimate continuum methodology differences rather than a bad fit, and the $\sim$28\% false-negative rate is consistent with the tail of a distribution whose median is 5--7~m\AA. The flagging system is designed to remove catastrophic failures, not to minimize disagreement with any particular reference.}

\subsection{Comparison to Expert Measurements}
\label{sec:comparison}

Figure~\ref{fig:onetoone} compares Egent measurements against C3PO values for representative Magellan/MIKE spectra, showing both GPT-5 (top row) and GPT-5-mini (bottom row) results. \rev{For each spectrum, we compute a per-spectrum $R^2$ from the ordinary least-squares linear regression of Egent EW versus C3PO EW, using only unflagged (accepted) lines and requiring $\geq$20 matched lines per spectrum. Spectra with $R^2 < 0.8$ are excluded as a quality gate, since low $R^2$ indicates a systematic mismatch (e.g., from wavelength calibration errors or data reduction issues) rather than line-by-line scatter. In practice, all 21 GPT-5 spectra and all 63 GPT-5-mini spectra pass this threshold, as the input sample was pre-selected for reliable wavelength calibration before processing.} Figure~\ref{fig:r2_slope} summarizes the distribution across all spectra with C3PO matches: both models achieve $R^2 > 0.82$ (median $\approx$ 0.93), with slopes scattered around unity (median $\approx$ 1.03). \rev{Egent produces measurements with a raw median absolute deviation (MAD) of 5--7~m\AA\ ($\sim$11--12\% of mean EW) against expert values, with no post-hoc per-spectrum correction applied}---quality that traditionally requires substantial human effort.

\begin{figure*}[t]
    \centering
    \includegraphics[width=\textwidth]{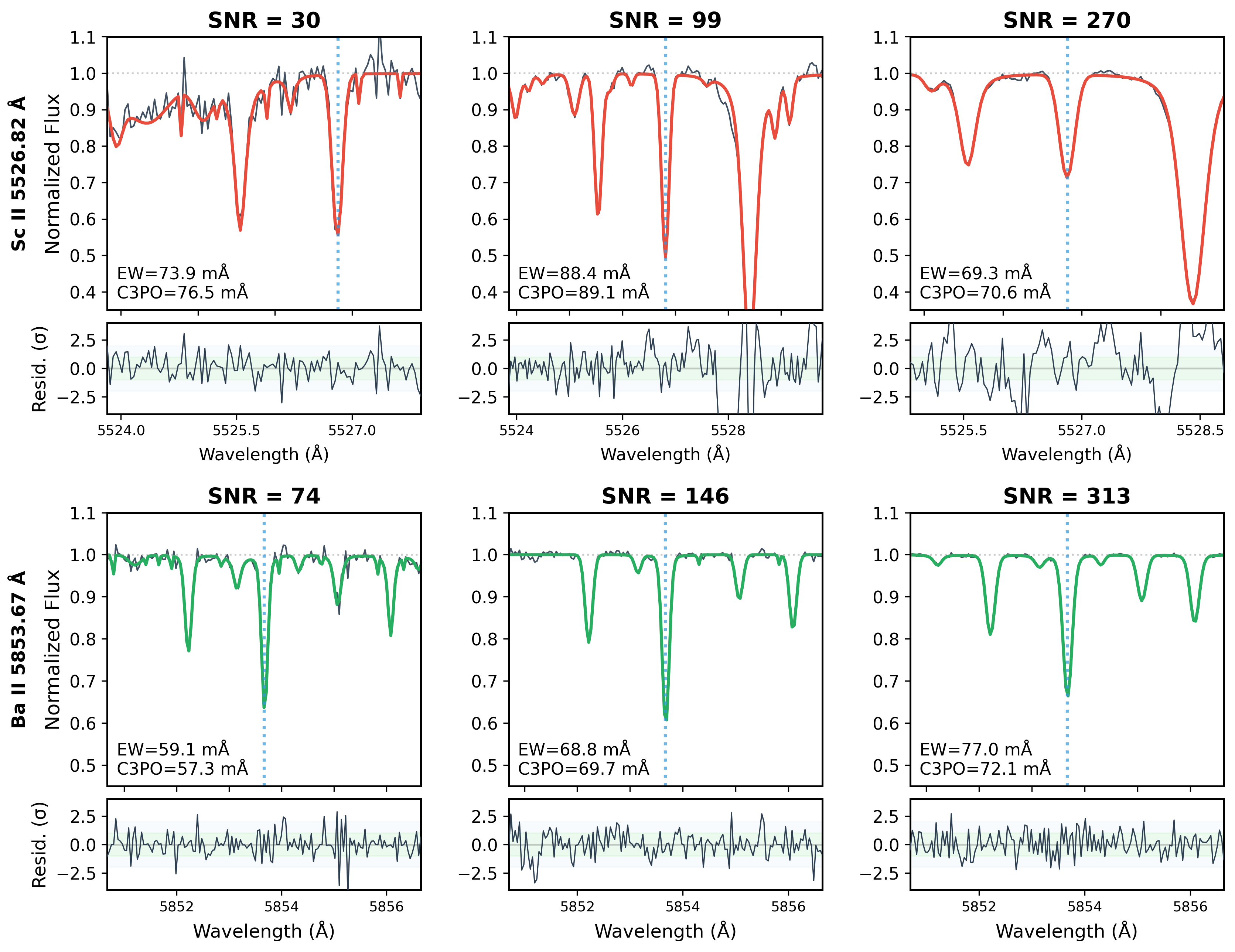}
    \caption{Fitting quality across SNR levels for two representative lines: Sc~II 5526.82~\AA\ (red, top) and Ba~II 5853.67~\AA\ (green, bottom). Each column shows the same line observed in different spectra at SNR~=~30--313. Top panels: normalized flux (dark) with Voigt model overlay. Bottom panels: normalized residuals with $\pm1\sigma$ (green) and $\pm2\sigma$ (blue) bands. Even at SNR~=~30, the measured EW agrees with C3PO to within 3~m\AA. Residuals scale appropriately with noise level without systematic structure.}
    \label{fig:snr_comparison}
\end{figure*}

Despite the strong correlation, Figure~\ref{fig:onetoone} reveals clear spectrum-to-spectrum slope offsets: GPT-5 panels show slopes of 0.93--0.97, while GPT-5-mini panels range from 0.96--1.04. These offsets arise from continuum methodology differences, not fitting failures---visual inspection of individual fits (Appendix~D) confirms that Egent's Voigt profiles trace the data precisely. In traditional manual analysis, the common practice is to first globally normalize the spectrum before local continuum adjustment---a step that facilitates human inspection but introduces systematic choices that propagate to the final EW. C3PO uses such pre-normalized spectra, while Egent fits directly from raw flux with no pre-normalization step. The slope distribution width ($\pm$15\%) therefore reflects the inherent subjectivity of global continuum placement. In the linear regime of the curve of growth where EW $\propto N$, a slope deviation translates directly to abundance error via $\Delta$[X/H] = log$_{10}$(slope)---e.g., slopes of 0.93 and 1.04 correspond to $-0.03$ and $+0.02$~dex systematic offsets, comparable to or larger than random measurement precision. Our continuum-less approach explicitly avoids this source of systematic error.

Table~\ref{tab:llm_stats} shows that GPT-5-mini achieves a higher acceptance rate (88.9\% versus 80.9\% for GPT-5), suggesting GPT-5 is more conservative in its flagging. Both models produce high-quality measurements on accepted lines, with comparable $R^2$ and MAD values. The choice between models depends on the desired balance between throughput and stringency: GPT-5-mini offers 5$\times$ cost savings with slightly higher acceptance rates, while GPT-5 may provide more conservative quality control. That modern AI, even a lightweight vision model, can achieve expert-level quality at $\sim$200 lines per dollar represents a fundamental shift in the cost-benefit calculus for line-by-line spectroscopic analysis.

Figure~\ref{fig:deviation_by_flag} demonstrates that AI flagging effectively identifies problematic measurements: accepted lines (green) cluster tightly at small deviations from C3PO, while flagged lines (red) show broader distributions extending to larger errors. In Figure~\ref{fig:onetoone}, this manifests as hollow red markers clustering at larger deviations from the 1:1 line. However, some unflagged points also show large deviations, and some flagged lines have small deviations---visual inspection (Appendix~D) reveals that occasional errors escape flagging, just as human measurements are not immune to error, while some valid flagging may catch fits that happen to land near catalog values by chance. One could impose more stringent flagging criteria, but this trades completeness for purity; the optimal threshold depends on the scientific use case. \rev{A sensitivity analysis confirms this trade-off is benign: varying the RMS flagging threshold from a strict 2.0$\sigma$ to a permissive 5.0$\sigma$ changes the acceptance rate from 72\% to 91\%, while the MAD shifts by only 0.2~m\AA\ (5.5 to 5.7~m\AA). The default 3.0$\sigma$ threshold provides a reasonable balance, and users can adjust it with minimal impact on measurement accuracy.} The key advantage is that AI flagging provides reproducible and homogeneous quality control---a form of systematic reasoning that can be version-controlled and applied uniformly across large datasets.

\rev{An important practical question is whether the reported $\sigma_{\rm EW}$---propagated from the \texttt{curve\_fit} covariance matrix---is well-calibrated. We assess this by comparing $\sigma_{\rm EW}$ against $|\Delta\mathrm{EW}|$ residuals with C3PO, noting at the outset that C3PO measurements are not ground truth: they carry their own uncertainties from manual continuum placement and line-profile judgments. The observed scatter therefore reflects contributions from both pipelines, and any miscalibration factors derived below are upper bounds on Egent's uncertainty alone.}

\rev{With this caveat, we find the reported uncertainties underestimate the joint scatter by a factor of $\sim$2--3 overall: for GPT-5-mini, only 32\% of deviations fall within $1\sigma_{\rm EW}$ (versus 68\% expected for a Gaussian). The miscalibration is strongly regime-dependent. Lines with small reported $\sigma_{\rm EW} < 2$~m\AA\ show the largest discrepancy (ratio $\approx$~3.6), because these are high-SNR clean lines where photon noise is well-captured but the dominant error source---continuum methodology differences between the two pipelines---is not. In contrast, lines with large reported $\sigma_{\rm EW} \geq 5$~m\AA\ are well-calibrated (ratio $\approx$~1.0), because the fitting uncertainty itself is large enough to encompass the systematic floor. The reported $\sigma_{\rm EW}$ should therefore be interpreted as a lower bound capturing statistical fitting uncertainty only. For applications requiring conservative total uncertainties, one may add a systematic floor in quadrature: $\sigma_{\rm total} = \sqrt{\sigma_{\rm EW}^2 + \sigma_{\rm sys}^2}$, with $\sigma_{\rm sys} \leq 5$~m\AA\ as a safe upper bound---the true Egent systematic floor is likely smaller, since part of the observed scatter originates in the reference measurements.}

\rev{To place these results in the context of established automated EW codes: ARES \citep{Sousa2007, Sousa2015} and DAOSPEC \citep{Stetson2008} are the most widely used alternatives for batch EW measurement. A direct head-to-head comparison on the same spectra is not straightforward because both codes require pre-normalized input spectra, whereas Egent operates on raw flux; applying a global normalization to our data would introduce continuum choices that confound the comparison. We therefore compare against published scatter figures. On pre-normalized synthetic solar spectra ($R \sim 65{,}000$, SNR~$\sim$~100), ARES achieves rms scatter $<$~2~m\AA\ against manual \texttt{splot} measurements, with a systematic offset below 1.5~m\AA\ \citep{Sousa2007}. On real FEROS spectra (SNR~$>$~200, $R \sim 50{,}000$), the mean ARES--manual difference remains $\sim$1.5~m\AA. When comparing ARES~v2's fully automatic continuum mode against manually tuned settings on the same code, \citet{Sousa2015} find dispersion $\sim$4.4~m\AA\ across $\sim$380 lines per spectrum on seven spectra spanning SNR~70--2000. For DAOSPEC, \citet{Stetson2008} report scatter of 3.7~m\AA\ against ARES on a high-SNR Solar HARPS spectrum ($R \sim 45{,}000$, SNR~$\sim$~350, 98 lines), rising to 10.3~m\AA\ on moderate-SNR spectra of $\omega$~Centauri red giants ($R \sim 45{,}000$, SNR~$\sim$~100--150, 1150 lines across six stars). Egent's MAD of 5--7~m\AA\ against an independent reference pipeline is competitive with these benchmarks, particularly given that (1)~it is measured against a separate pipeline rather than the same code with different settings, (2)~it includes the full range of line strengths and spectral conditions without pre-selection, and (3)~it requires no pre-normalized spectra. Egent's direct-fit pathway---without LLM intervention---already achieves MAD~=~5.4~m\AA\ on accepted lines, demonstrating that the core fitting engine performs at the level of established tools; the LLM adds completeness and provenance rather than raw precision.} \revtwo{We emphasize that the comparison above relies on published scatter figures rather than a head-to-head run on identical inputs; a controlled benchmark---running ARES and DAOSPEC on the same C3PO spectra after a normalization step matched to each code's input requirements, with results stratified by line strength and line density---would give a more definitive answer on the relative value of Egent versus established automated EW pipelines, and is a natural future extension.}

\subsection{Performance Across SNR}
\label{sec:snr_test}

Rather than artificially degrading spectra, we assess SNR robustness using the natural variation in our GPT-5-mini sample. The 63 spectra span median SNR from 50 to 250 per pixel, providing a direct test of performance across data quality regimes. Table~\ref{tab:snr_spectrum} shows that correlation quality ($R^2 \geq 0.90$, median 0.93) and precision (MAD $\sim$ 4--8~m\AA) remain stable across all SNR bins.

\begin{table}[h]
\centering
\caption{Per-Spectrum Statistics by SNR (GPT-5-mini)}
\label{tab:snr_spectrum}
\begin{tabular}{lccc}
\hline
SNR Range & N Spectra & $R^2$ (median) & MAD (median) \\
\hline
$<$100 & 7 & 0.90 & 5.7~m\AA \\
100--150 & 12 & 0.95 & 3.5~m\AA \\
150--200 & 32 & 0.93 & 8.0~m\AA \\
$>$200 & 12 & 0.93 & 4.2~m\AA \\
\hline
All & 63 & 0.93 & 5.7~m\AA \\
\hline
\end{tabular}
\end{table}

Figure~\ref{fig:snr_comparison} illustrates this robustness with two representative lines---Sc~II 5526.82~\AA\ and Ba~II 5853.67~\AA---observed at SNR ranging from 30 to 313. Even at SNR~=~30, where individual pixel fluctuations reach 3\% of the continuum, the Voigt fit successfully captures the line profile and yields EW within 3~m\AA\ of the expert value. The residuals naturally show larger scatter at low SNR, but remain within the expected noise envelope without systematic structure.

To disentangle the effects of SNR and line strength, Tables~\ref{tab:flag_snr} and \ref{tab:mad_snr} present flagging rate and MAD as two-dimensional functions of both variables. The practical lower limit for reliable measurement depends on line strength: typical lines (EW~$\sim$~50--100~m\AA) remain robust down to SNR~$\sim$~30--50, while weaker lines (EW~$<$~20~m\AA) require higher SNR to distinguish from noise.

\begin{table}[h]
\centering
\caption{Flagging Rate (\%) by SNR and Catalog EW}
\label{tab:flag_snr}
\begin{tabular}{l|ccccc|c}
\hline
 & \multicolumn{5}{c|}{SNR} & \\
EW (m\AA) & $<$100 & 100--150 & 150--200 & 200--300 & $>$300 & All \\
\hline
$<$30 & 5 & 3 & 4 & 7 & 5 & 5 \\
30--60 & 7 & 5 & 7 & 7 & 6 & 7 \\
60--100 & 5 & 9 & 9 & 9 & 8 & 8 \\
100--150 & 12 & 9 & 10 & 12 & 11 & 11 \\
\hline
\end{tabular}
\end{table}

\begin{table}[h]
\centering
\caption{MAD (m\AA) by SNR and Catalog EW (unflagged lines)}
\label{tab:mad_snr}
\begin{tabular}{l|ccccc|c}
\hline
 & \multicolumn{5}{c|}{SNR} & \\
EW (m\AA) & $<$100 & 100--150 & 150--200 & 200--300 & $>$300 & All \\
\hline
$<$30 & 5.2 & 3.7 & 3.7 & 3.7 & 4.0 & 3.9 \\
30--60 & 7.0 & 5.0 & 5.3 & 5.7 & 6.6 & 5.9 \\
60--100 & 8.8 & 7.3 & 8.2 & 7.5 & 9.1 & 8.1 \\
100--150 & 11.6 & 11.1 & 10.6 & 9.7 & 10.1 & 10.4 \\
\hline
\end{tabular}
\end{table}

The flagging rate (Table~\ref{tab:flag_snr}) is remarkably uniform across SNR (5--12\% in all bins), indicating that quality control responds to fit quality rather than raw noise level. The modest increase with line strength (5\% for weak lines to 11\% for strong lines) likely reflects higher blend probability in crowded spectral regions.

The MAD values (Table~\ref{tab:mad_snr}) tell a similar story: measurement precision is largely independent of SNR. At low SNR ($<$100), the MAD ranges from 5.2~m\AA\ for weak lines to 11.6~m\AA\ for strong lines---nearly identical to the 4.0--10.1~m\AA\ range at high SNR ($>$300). \rev{This flatness is expected given the uncertainty calibration results above: the systematic floor from continuum methodology differences ($\leq$5~m\AA) dominates over photon noise across the full SNR range of our sample. At SNR~=~100, the photon-noise contribution to a typical line (EW~$\sim$~60~m\AA\ spanning $\sim$30 pixels) is $\sim$2--3~m\AA, below the systematic floor. A noise-injection experiment would therefore primarily confirm what the flat MAD already demonstrates---that our sample is in the systematics-dominated regime. Reaching photon-noise-dominated conditions for typical lines would require SNR~$\lesssim$~20--30, below which continuum fitting itself becomes unreliable.} The MAD increases with line strength in absolute terms (4--10~m\AA) but decreases in relative terms (22\% for weak lines to 9\% for strong lines)---the expected behavior where absolute uncertainty is roughly constant while relative precision improves for stronger features.

\section{Discussion}
\label{sec:discussion}

This work demonstrates that an autonomous LLM agent, serving as both validator and adjuster, can perform EW measurement at human expert level. The key is providing the agent with appropriate tools and letting it reason about fits visually---the same process a human expert would follow, but automated and reproducible. Several implications and caveats warrant discussion.

\subsection{Implications for Surveys}

The ability to measure EWs automatically, reproducibly, and with full provenance has implications for spectroscopic surveys. Currently, even high-resolution surveys rely on full-spectrum fitting partly because manual EW measurement cannot scale to hundreds of thousands of stars. With Egent-like pipelines, line-by-line analysis becomes feasible at survey scale---at least for resolved lines at $R > 20,000$.

A key contribution of this work is demonstrating a complete end-to-end pipeline: starting from reduced flux spectra, everything proceeds autonomously to produce an EW table where every fit is visualized and flagged. No intermediate steps require human intervention.

The computational requirements are modest. Direct Voigt fitting requires $\sim$0.1 seconds per line. LLM inspection adds latency---GPT-5 requires $\sim$1 second per iteration (up to 10 iterations for difficult cases), while GPT-5-mini operates at sub-second speeds. The process is massively parallelizable, limited primarily by CPU cores (for generating diagnostic PNG plots, which proves more expensive than the fitting itself) and API rate limits. In our GPT-5-mini runs, we sustained 80 concurrent workers without hitting rate limits; the bottleneck was local plot generation, easily addressed with a small computing cluster.

To estimate survey-scale feasibility: for the $\sim$200-line lists typical of differential abundance studies, at $\sim$1 second per line, a single spectrum completes in $\sim$3 minutes. With 1000 parallel workers, 500,000 spectra (comparable to APOGEE) would complete in $\sim$2 days of wall-clock time. For analyses requiring more lines per spectrum, computation time scales proportionally.

The scientific potential is considerable. As discussed in the introduction, EW-based differential analysis persists because it can cancel systematic errors that plague full-spectrum fitting. Formulating proper differential measurements that harness millions of spectra with billions of individual EW measurements represents a scale not yet explored---but one that autonomous pipelines now make accessible.

Current differential abundance studies use carefully curated line lists containing $\sim$200 lines chosen for minimal blending and reliable atomic data. We use these lines because they are easy to measure reliably, and we measure them reliably because experts have accumulated experience with them. If all lines could be measured automatically, new approaches to abundance determination become possible. Self-calibration methods could use the full line list to constrain both abundances and atomic data simultaneously.

\subsection{Reproducibility and Human-in-the-Loop Development}

Every Egent measurement includes complete provenance: Voigt parameters, continuum coefficients, quality metrics, and (when triggered) the full LLM conversation. This differs qualitatively from manual measurement, where a published EW cannot be traced to specific decisions about continuum placement or blend handling.

A strength of the agentic approach is that human expertise enters through prompts and tool definitions, which can be iteratively refined. Much of Egent's development involved examining cases where initial prompts failed, understanding why, and improving the instructions. We employed an outer loop where an LLM evaluated failure cases on small test sets, iterating over different prompt formulations---effectively using AI to help develop AI prompts. This process led to the final prompts presented in Appendix~\ref{app:prompt}.

For example, early versions of the prompt did not explicitly mention W-shaped residuals as indicators of blends; adding this guidance improved blend detection. Users can also incorporate domain knowledge---for instance, flagging specific wavelength regions with known blends that warrant extra attention. This refinement process is more systematic than the implicit expertise accumulated in a human measurer's head---and it is version-controlled. Once prompt development is complete, everything proceeds autonomously. The agent mimics the same iterative inspection process that a human expert would perform, but without the tedium that makes manual measurement impractical at scale.

\rev{It is useful to distinguish two forms of reproducibility that the pipeline provides. The first is \textit{archival reconstruction}: every completed run stores the full Voigt parameters, continuum coefficients, and LLM reasoning chain, so any measurement can be exactly reconstructed from the saved JSON without re-running the model. This form of reproducibility is deterministic and permanent. The second is \textit{prospective reprocessing}: re-running the pipeline on the same data at a later date. Because LLM outputs depend on model version, sampling temperature, and stochastic decoding, re-processing will not produce bit-identical results.}

\rev{We quantify this variation using the cross-model comparison as an upper bound, since GPT-5 and GPT-5-mini are architecturally different models. Among 3,640 lines measured by both models on the same spectra: 55\% have exactly identical EW (both accepted the deterministic direct fit), 73\% agree within 1~m\AA, and 96\% agree within 10~m\AA. The 90th-percentile difference is 5.0~m\AA\ with no systematic bias (mean $\Delta$EW~$=$~$-0.07$~m\AA). Even for lines where both models invoked the LLM (n~$=$~2,632), the median difference is only 0.3~m\AA. This cross-model scatter is comparable to the systematic floor from continuum methodology, indicating that model choice is not the dominant source of uncertainty. Run-to-run variability with the same model would be smaller still. For long-term scientific use, we recommend archiving the stored outputs rather than relying on re-execution, and recording the model version used for each run.} \revtwo{The cross-model comparison reported here is an upper bound rather than a direct measurement of stochastic variability; a dedicated same-model repeatability experiment---re-running identical prompts at fixed sampling temperature on the same lines and the same model version---would provide a more definitive lower bound on the run-to-run noise floor under model drift, and is a natural future test.}

\subsection{Offline and Web Interface Versions}

While the core pipeline is compact---a few hundred lines of Python with API calls---we recognize that deploying commercial LLM APIs may present barriers for some users. We therefore provide two additional deployment options.

First, we offer a fully offline version using Qwen-3-VL-8B \citep{Yang2025}, an open-source vision-language model deployable through the Ollama ecosystem. The entire installation is self-contained and runs on consumer hardware (e.g., a Mac with default RAM). This is particularly valuable for observatory settings where external internet access through APIs may raise security concerns. While processing a few spectra is feasible on such hardware, speed is limited without GPU acceleration---generating reasoning tokens benefits from dedicated GPU hardware. We do not provide formal benchmarks as we lack a cluster of Macbooks for systematic testing. However, given that GPT-5-mini performs well, and lightweight open-source vision-language models have been reported to achieve comparable (or superior) performance at the small-model end, we expect the offline version to produce reliable results, albeit more slowly. 

Second, we provide a web interface for drag-and-drop analysis. Users can upload spectra without requiring their own API key---the web service covers the LLM costs. An initial preprocessing step converts arbitrary formats into Egent's expected input. While this interface does not support parallel processing, it enables quick testing and demonstration without any local installation or API configuration. 

Installation instructions for both the Ollama backend and the web interface are available at \url{https://github.com/tingyuansen/Egent}.

\subsection{Limitations}

The monetary cost of LLM APIs represents a practical limitation. For GPT-5-mini, processing costs approximately \$0.005 per line ($\sim$200 lines per dollar); GPT-5 is $\sim$5$\times$ more expensive ($\sim$40 lines per dollar). A full spectrum of 200 lines costs approximately \$1 for GPT-5-mini. For an APOGEE-scale survey (500,000 spectra $\times$ 200 lines = 100 million lines), total API cost would be $\sim$\$500,000 for GPT-5-mini---not trivially cheap, but not impossible for a major survey either. 

These estimates represent an upper limit. One straightforward way to reduce costs is relaxing the quality thresholds that trigger LLM inspection, relying more heavily on algorithmic checks and reserving visual inspection for the most uncertain cases. In our current configuration, $\sim$60--65\% of lines trigger LLM inspection; tightening acceptance criteria for direct fits could reduce this fraction substantially. Fine-tuned lightweight open-source models (1--2B parameters) represent another promising direction: if domain-specific fine-tuning can achieve comparable performance to general-purpose frontier models, costs could decrease by orders of magnitude while enabling fully offline operation. As LLM capabilities continue to improve and costs continue to decline, these estimates will become increasingly conservative.

Another limitation is uncertainty about how Egent scales to lower resolution or SNR. Our validation uses C3PO spectra, which by survey design represent the high-SNR regime (SNR~$>$~50). Egent is designed for high-resolution ($R > 20,000$), high-SNR spectra of solar-type stars. At lower resolution, blending becomes severe and the ``local continuum'' approximation breaks down---though this is not a fundamental limitation of the agentic approach, but rather would require more sophisticated priors (e.g., from spectral synthesis). Testing has focused on solar-type stars; extension to metal-poor stars (where lines are weaker and sparser), rapid rotators (where lines are broadened and blended), and emission-line objects would require validation. The framework should generalize, but prompts and quality thresholds may need adjustment.

\rev{Regarding line profiles: the Voigt model is adequate for the slowly rotating ($v\sin i \lesssim 5$~km\,s$^{-1}$), solar-type stars in our validation sample, where macroturbulent and rotational broadening are well approximated as Gaussian contributions absorbed into the $\sigma$ parameter. For rapid rotators or very high resolution spectra where asymmetric broadening kernels become discernible, explicit rotational or radial-tangential macroturbulence profiles would be needed; this is a scope limit of the current implementation, not of the agentic framework. Simultaneous fitting of continuum and multiple Voigt components in narrow windows can become weakly constrained in crowded regions; in practice, the LLM-driven iterative refinement (adjusting windows, continuum order, and component count) serves as a practical stability diagnostic, and fits that remain unstable are flagged. Because the entire fitting engine is written from scratch in pure Python with minimal dependencies (\texttt{scipy}, \texttt{numpy}), adding new profile functions (e.g., a rotational kernel or radial-tangential macroturbulence) requires only implementing the profile and registering it as a tool---no modification of legacy compiled code is needed.}

\rev{The fitting engine uses standard Levenberg--Marquardt least squares rather than robust alternatives such as IRLS with a Huber or Tukey loss. Analysis of the quality issues that trigger LLM intervention reveals that the dominant failure modes are correlated residuals from missed blends (present in $>$80\% of triggered fits) and wrong continuum shape---both structural model-mismatch problems that robust regression would not resolve, since the correct response is adding a Voigt component or changing the continuum model, not downweighting individual pixels. The continuum fitting already incorporates iterative $2.5\sigma$ clipping, which provides robustness against absorption features during initial continuum estimation. Robust regression and LLM inspection thus target orthogonal failure modes: the former addresses point-outlier sensitivity, while the latter handles model selection (which components, which continuum, which window). The two are complementary, and adding a robust loss function to the fitting engine is a straightforward future extension.} \revtwo{The argument above is based on inspection of the failure-mode statistics rather than a side-by-side ablation; a controlled deterministic test---e.g., swapping the standard Levenberg--Marquardt loss for an IRLS scheme with a Huber or Tukey biweight loss within the same engine, and then re-measuring how often the LLM is still triggered and whether accepted-line scatter changes---would more definitively quantify what fraction of the inspection benefit is attributable to robust statistics versus model-guided adjustments.}

Egent addresses the labor-intensive bottleneck in EW measurement, but it is not a complete end-to-end abundance pipeline. Other steps---such as rest-frame velocity correction via cross-correlation---remain outside its scope, though these are largely programmable tasks that do not require the same human judgment. The natural extension is connecting EW measurement to abundance determination via curve-of-growth analysis or spectral synthesis. These steps are computationally inexpensive and well-understood; keeping them separate maintains modularity and avoids entangling measurement uncertainty with model uncertainty. Indeed, with reliable EWs in hand, many spectral models can be tested rapidly.

Perhaps the most fundamental limitation lies in current LLM capabilities, particularly visual reasoning. While LLM inspection helps with flagging and edge cases, we have seen instances where LLM reasoning produces worse fits than direct fitting alone, and occasionally fails to flag genuinely bad fits. The pipeline is not foolproof---but neither are human experts. We also observe that current LLMs are not particularly proficient at function tool use, contributing to the $\sim$10\% flagging rate even at high resolution. A natural path forward would be fine-tuning models with reinforcement learning (e.g., GRPO), which does not require labeled data, but this is beyond the scope of this paper.

\subsection{Broader Context}

LLM agents have generated considerable excitement, with a rush to tackle ambitious problems like automated hypothesis generation. Yet even in this relatively simple case---eight tools, a well-defined task, data the models were unlikely trained on---out-of-the-box models without human-in-the-loop refinement prove challenging. While visual reasoning is strong and enables effective flagging, reliable autonomous operation required careful prompt engineering and constraint building.

This is not to say that agents cannot drastically expedite research. This work demonstrates multiple layers of agentic improvement: using agents to refine prompts under human supervision, building in domain constraints, and iterating on failure cases. For tasks that require serious repetition yet demand judgment---exactly the characteristics of EW measurement---the agentic framework provides a successful model.

The approach generalizes to other astronomical tasks with similar characteristics: damped Lyman-$\alpha$ identification, spectral classification, and other problems where programmatic rules fail on edge cases but human judgment succeeds. As LLMs strengthen, particularly in visual reasoning and tool use, this represents a growing avenue for automating traditionally manual analysis.

\section{Conclusion}
\label{sec:conclusion}

We present Egent, an autonomous agent for equivalent width measurement that combines classical multi-Voigt profile fitting with LLM-based visual inspection. The system operates directly on raw flux spectra without requiring pre-normalized continua, and provides complete provenance for every measurement.

\rev{Validation against 18,615 lines from 84 Magellan/MIKE spectra measured by human experts in the C3PO program yields a raw MAD~=~5--7~m\AA\ agreement without any post-hoc per-spectrum correction. Per-spectrum slopes scatter within $\sim$15\% of unity (median $\approx$ 1.03), reflecting differences in global continuum methodology between Egent's local fitting and traditional approaches rather than fitting failures; the overall correlation is strong ($R^2 \approx 0.93$).} The pipeline is robust across SNR$\sim$50--250, with MAD increasing only modestly from $\sim$5~m\AA\ at high SNR to $\sim$8~m\AA\ at SNR~$<$~75.

The LLM's primary role is quality control rather than systematic improvement. Approximately 60--65\% of lines trigger LLM inspection and are subsequently refined; the LLM confirms good fits, flags problematic cases ($\sim$10--20\%), and occasionally rescues edge cases through tool use. Agreement between GPT-5 and GPT-5-mini demonstrates reproducibility and validates that the pipeline does not depend on model-specific behaviors.

The computational requirements are practical. Direct fitting requires $\sim$0.1 seconds per line; LLM inspection adds $\sim$1 second per iteration. A typical 200-line spectrum costs $\sim$\$1 for GPT-5-mini; survey-scale processing is feasible but not trivially cheap, motivating future work on fine-tuned lightweight models. We provide three deployment options: API-based processing for production runs, a fully offline version using Qwen-3-VL-8B through Ollama for secure environments, and a web interface for drag-and-drop analysis without local installation.

Every measurement stores complete Voigt parameters, continuum coefficients, quality metrics, and (when triggered) the full LLM reasoning chain. This provenance enables exact reconstruction of any fit without re-running the analysis---a qualitative improvement over traditional manual measurement where decisions are rarely recorded.

Egent demonstrates that LLM agents can perform expert-level judgment tasks in astronomy when provided with appropriate tools and domain constraints. The approach generalizes to other problems where programmatic rules fail on edge cases but human judgment succeeds. By removing the human bottleneck while maintaining reliability and transparency, Egent opens the path to line-by-line abundance analysis at survey scale---enabling differential analysis methods that harness millions of spectra with billions of individual EW measurements.

\acknowledgments

This project was originally inspired by the first author's teaching of Astron 1221 (Data Analysis in Astronomy) at OSU, a first-year Python course for astronomers covering both spectroscopic data analysis and basic agentic AI frameworks---which the author deems important even for freshman students to thrive in this era. The pipeline was initially conceived as a class project demonstrating how LLM function tools could be applied to a real astronomical problem; since freshman students tend to gravitate toward either amateur astronomy problems (lunar phases, asteroid impacts) or Nobel Prize-winning topics (S-stars near Sgr~A$^*$, 51~Pegasi~b)---while all using similar tools---the author decided to take on this project himself.

The recent spectroscopy workshop held by Catherine Manea and Madeline McKenzie inspired the author to finish this project. We appreciate fruitful discussions with David Weinberg, Jennifer Johnson, Lucy Lu, Tawny Sit, Sharon Wang, Qinghui Sun, and Megan Bedell throughout the year. Part of the routine is inspired by the ongoing PASTA project (similar to C3PO), and the author's lack of funding to hire an expert for yet another painstaking EW measurement campaign---combined with OSU's recently grown fortuitous LBT time fraction, which will lead to more PEPSI spectra that will only accentuate this problem.

This research was supported by NSF Grant AST-2406729 and a Humboldt Research Award from the Alexander von Humboldt Foundation. The GPT-5 and GPT-5-mini API allocation in this development was substantial and is supported by the NSF NAIRR program. The authors acknowledge the National Artificial Intelligence Research Resource (NAIRR) Pilot and OpenAI for contributing to this research result.

While the ideas, draft, and layout are all from the authors, this project has benefited---in line with its own ideals---from collaboration with Claude-Opus-4.5 and Cursor for both coding development and writing refinement. Google's Nanobanana-Pro is used to generate the schema in Figure 2.

We thank D.~Yong's right hand for the C3PO measurements that enabled validation.

\appendix

\section{System Prompts}
\label{app:prompt}

We provide the complete system prompt used for LLM visual inspection, enabling full reproducibility.

\subsection{Visual Inspection Prompt}

The following prompt instructs the LLM on its role as an expert visual reviewer for borderline EW measurements. When the direct multi-Voigt fit produces quality metrics outside acceptable thresholds (Section~\ref{sec:quality}), the LLM is invoked to inspect the diagnostic plot and decide whether to accept, improve, or flag the measurement. A key feature of this prompt is the explicit instruction to focus on the target region ($\pm$0.5~\AA) and ignore edge effects that do not affect the measurement.

\begin{tcolorbox}[colback=black!3, colframe=black!60, fonttitle=\bfseries, breakable]

You are an expert reviewer for stellar EW measurements. You're called for BORDERLINE cases where automated fitting is uncertain.
\vspace{1em}

\textbf{YOUR ROLE:} Use visual inspection to make judgment calls that algorithms can't.

\vspace{1em}

\textbf{WORKFLOW:} (1) The fit is already done. Call get\_fit\_plot to VISUALLY INSPECT. (2) Make a judgment: ACCEPT, IMPROVE, or FLAG.

\vspace{1em}

\textbf{$\Rightarrow$ CRITICAL: FOCUS ONLY ON TARGET REGION $\Leftarrow$}

The TARGET LINE is marked with a BLUE DASHED vertical line. ONLY care about fit quality within $\pm$0.5~\AA\ of this target.

\textbf{EDGE vs TARGET distinction:}
\begin{itemize}[noitemsep,topsep=0pt]
\item IGNORE: Spikes/noise at window edges ($>$1.5~\AA\ from target)---these are normal
\item CHECK: Residuals within $\pm$0.5~\AA\ of the blue target line
\item If target region is clean but edges are messy $\rightarrow$ ACCEPT
\item If target region has W-pattern or blend $\rightarrow$ FIX or FLAG
\end{itemize}

\vspace{1em}

\textbf{RMS INTERPRETATION:}
\begin{itemize}[noitemsep,topsep=0pt]
\item \textbf{RMS $<$ 1.5$\sigma$:} LIKELY GOOD---still check for W-shaped residuals at target
\item \textbf{RMS 1.5--2.0$\sigma$:} GOOD---check target region only, accept if clean
\item \textbf{RMS 2.0--2.5$\sigma$:} MARGINAL---try to improve, accept if target is clean after
\item \textbf{RMS $>$ 2.5$\sigma$:} POOR---needs work or FLAG
\end{itemize}

\vspace{1em}

\textbf{BLEND DETECTION (CHECK EVEN WITH LOW RMS):}

Look for these patterns in residuals NEAR THE TARGET ($\pm$1~\AA):
\begin{itemize}[noitemsep,topsep=0pt]
\item \textbf{W-SHAPED residuals} at target = MISSED BLEND. Fix: fit\_ew(additional\_peaks=[wavelength])
\item \textbf{ASYMMETRIC wing/shoulder} on target line. Fix: add the wing as additional\_peak
\item \textbf{Systematic offset} at target center. May need continuum adjustment or nearby blend
\end{itemize}

\vspace{1em}

\textbf{ACCEPT} (call record\_measurement) if: RMS $<$ 1.5$\sigma$ (excellent fit), OR RMS $<$ 2.0$\sigma$ AND target region residuals within $\pm$2$\sigma$.

\vspace{1em}

\textbf{IMPROVE} by trying: (1) Add missing peaks near target. (2) Polynomial continuum for curved residuals. (3) Window adjustment (narrower for edge issues, wider for truncated lines). (4) Manual continuum regions for crowded areas.

\vspace{1em}

\textbf{FLAG} (use sparingly!) only when: Problem is AT THE TARGET ($\pm$0.5~\AA), not at edges; RMS $>$ 2.5$\sigma$ that cannot be improved; target line is completely blended or absent.

\vspace{1em}

\textbf{DO NOT FLAG} for: Edge spikes/noise far from target ($>$1.5~\AA\ away); Low RMS ($<$2.0$\sigma$) with clean target region.

\end{tcolorbox}

\section{Tool Schemas}
\label{app:schemas}

Eight tools are available to the LLM agent, each implemented as an OpenAI function-calling schema. These tools mirror the operations a human spectroscopist would perform in IRAF or similar software: loading data, adjusting the wavelength window, configuring continuum fitting, adding blend components, inspecting results visually, and recording the final measurement.

\begin{tcolorbox}[colback=black!3, colframe=black!60, fonttitle=\bfseries, breakable]
\texttt{\textbf{load\_spectrum(gaia\_id: int)}} --- Loads a stellar spectrum from the data repository with barycentric correction applied. Returns metadata including SNR, wavelength range, and number of \'echelle orders available.\\[0.5em]

\texttt{\textbf{extract\_region(line\_wavelength: float, window: float = 3.0)}} --- Extracts the spectral region centered on the target wavelength. The \texttt{window} parameter specifies the half-width in \AA\ (default $\pm$3~\AA). The LLM uses this to ``zoom in'' (\texttt{window=2.0}) for isolated lines or ``zoom out'' (\texttt{window=5.0}) when strong lines are truncated at window edges.\\[0.5em]
\texttt{\textbf{set\_continuum\_method(method: str, order: int = 1)}} --- Configures the continuum fitting approach. Options: \texttt{iterative\_linear} (default), \texttt{iterative\_poly} (polynomial of specified order for curved continuum), \texttt{top\_percentile} (uses brightest pixels as anchors in crowded regions).\\[0.5em]
\texttt{\textbf{set\_continuum\_regions(regions: list)}} --- Last-resort tool for crowded regions where automatic continuum detection fails. The LLM visually identifies wavelength intervals that appear line-free and specifies them as \texttt{[[start1, end1], [start2, end2], ...]}. The continuum is then fit using only pixels within these regions.\\[0.5em]
\texttt{\textbf{fit\_ew(additional\_peaks: list = [])}} --- Performs the multi-Voigt fit using current extraction region and continuum settings. All catalog lines in the window are included automatically. The \texttt{additional\_peaks} parameter allows the LLM to specify wavelengths for Voigt components not in the catalog (for missed blends identified via W-shaped residuals). Returns EW, uncertainty, quality metrics, and full Voigt parameters for all fitted lines.\\[0.5em]
\texttt{\textbf{get\_fit\_plot()}} --- Generates the diagnostic plot for visual inspection. Returns a base64-encoded PNG showing: (1) normalized spectrum with model overlay, (2) individual Voigt component centers, (3) target wavelength indicator, and (4) normalized residuals with $\pm 1\sigma$ and $\pm 2\sigma$ bands.\\[0.5em]
\texttt{\textbf{flag\_line(line\_wavelength: float, reason: str)}} --- Flags a line as unreliable when no satisfactory fit can be achieved. Valid reasons: \texttt{no\_data}, \texttt{severe\_blend}, \texttt{fit\_failed}, \texttt{bad\_continuum}, \texttt{wrong\_line}. The flag and reason are stored in the output JSON.\\[0.5em]
\texttt{\textbf{record\_measurement(line\_wavelength: float, ew\_mA: float, ew\_err\_mA: float, quality: str)}} --- Records the final EW measurement with uncertainty and quality assessment (\texttt{good}, \texttt{acceptable}, or \texttt{poor}). This concludes the LLM inspection for accepted measurements.
\end{tcolorbox}

\section{Complete LLM Conversation Example}
\label{app:conversation}

Below we present a complete, verbatim conversation from GPT-5-mini processing Ca~I 6696.02~\AA\ (Gaia ID 55780840513067392). This case demonstrates blend detection via W-shaped residuals, peak injection to deblend, and the LLM's reasoning at each step. The initial direct fit triggered LLM inspection due to poor quality (central RMS = 3.63$\sigma$).

\begin{tcolorbox}[colback=black!3, colframe=black!60, title=\textbf{Iteration 1: Initial Visual Inspection}, fonttitle=\bfseries, breakable]
\textbf{LLM reasoning:} ``I will visually inspect the existing fit plot around 6696.02~\AA\ to check: (a) whether the red fit follows the black data at the target, (b) the residuals for any W-shaped pattern indicating missed blends (especially near 6696.0~\AA), and (c) whether the continuum shows a slope or offset. I'll call get\_fit\_plot to view the diagnostic.''\\[0.3em]
\textbf{Action:} \texttt{get\_fit\_plot(\{\})}
\end{tcolorbox}

\begin{tcolorbox}[colback=black!3, colframe=black!60, title=\textbf{Iteration 2: Blend Detection and Peak Injection}, fonttitle=\bfseries, breakable]
\textbf{LLM reasoning:} ``I observe a poor central fit: the central RMS is 3.63$\sigma$ ($\gg$2.5$\sigma$) and the residuals around 6696~\AA\ show a W-like pattern / asymmetric structure suggesting the target is blended (the fit treats the feature as a single symmetric component). In the main panel the data shows a slight asymmetry/shoulder near 6695.9--6696.1~\AA\ that the red fit doesn't fully reproduce. I will attempt to deblend by adding an additional Voigt component near 6695.9~\AA\ (keeping the main component at the target) to model the close blend.''\\[0.3em]
\textbf{Action:} \texttt{fit\_ew(\{"additional\_peaks": [6695.9]\})}
\end{tcolorbox}

\begin{tcolorbox}[colback=black!3, colframe=black!60, title=\textbf{Iteration 3: Re-inspection After Deblending}, fonttitle=\bfseries, breakable]
\textbf{LLM reasoning:} ``The updated fit deblended into multiple components but quality flags remain: correlated residuals and central RMS $>1.5\sigma$. I will inspect the updated diagnostic plot to locate remaining mismatches (any residual W-shapes or unmodeled dips around 6696.0--6696.3~\AA\ and at the right-side feature near 6698.65~\AA). If residuals show continuum tilt, I'll try a polynomial continuum next.''\\[0.3em]
\textbf{Action:} \texttt{get\_fit\_plot(\{\})}
\end{tcolorbox}

\begin{tcolorbox}[colback=black!3, colframe=black!60, title=\textbf{Iteration 4: Accept Measurement}, fonttitle=\bfseries, breakable]
\textbf{LLM reasoning:} ``After adding a deblending component at 6695.9~\AA\ the fit follows the data closely at the target. The residuals no longer show a W-shape at the target and the central RMS is 1.64$\sigma$ (overall RMS 1.49$\sigma$), which is within the acceptable range (1.5--2.5). There remains a neighboring fitted line at $\sim$6696.283~\AA\ that was modeled. I will record the measured EW from the improved fit.''\\[0.3em]
\textbf{Action:} \\
\texttt{record\_measurement(\{"line\_wavelength": 6696.02, "ew\_mA": 24.6, "ew\_err\_mA": 2.1, "quality": "acceptable"\})}
\end{tcolorbox}

\noindent This sequence---inspect, diagnose, adjust, re-inspect---mimics the workflow of a human expert but executes in seconds and produces a reproducible record. The explicit reasoning text documents \textit{why} each decision was made, not just what action was taken. The final EW (24.6~m\AA) differs from the initial direct fit (38.8~m\AA) by 37\%, demonstrating the importance of proper deblending.

\section{Individual Line Fits}
\label{app:linefits}

\begin{figure*}[p]
\centering
\includegraphics[width=\textwidth,height=0.92\textheight,keepaspectratio]{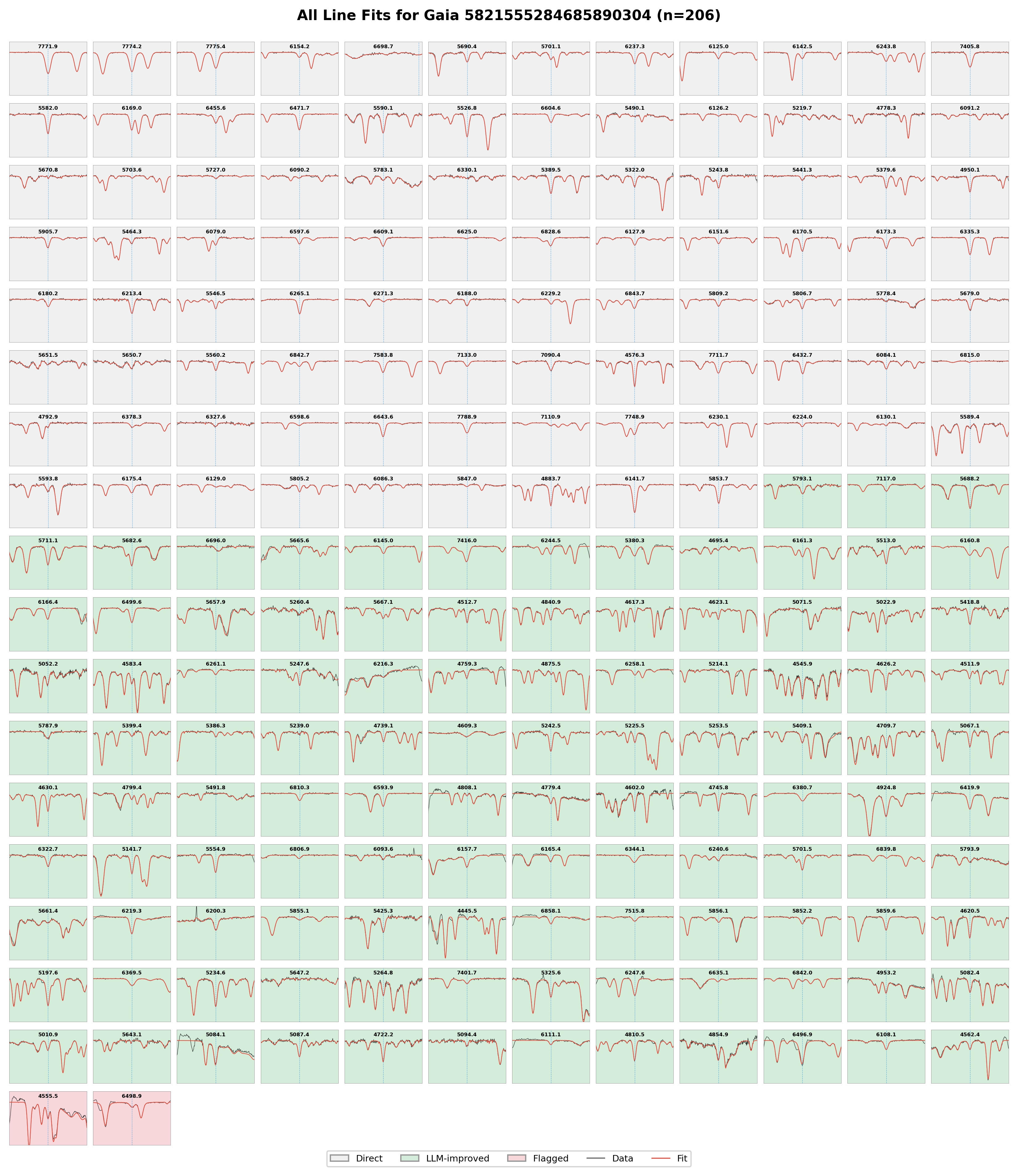}
\caption{All line fits for a representative spectrum. Each panel shows the target wavelength (\AA) above the plot, with normalized flux (black), multi-Voigt model (red), and target line position (blue dashed). Background colors: gray = direct fit accepted, green = LLM-improved, red = flagged. Lines are sorted by fit type.}
\label{fig:allfits1}
\end{figure*}

\begin{figure*}[p]
\centering
\includegraphics[width=\textwidth,height=0.92\textheight,keepaspectratio]{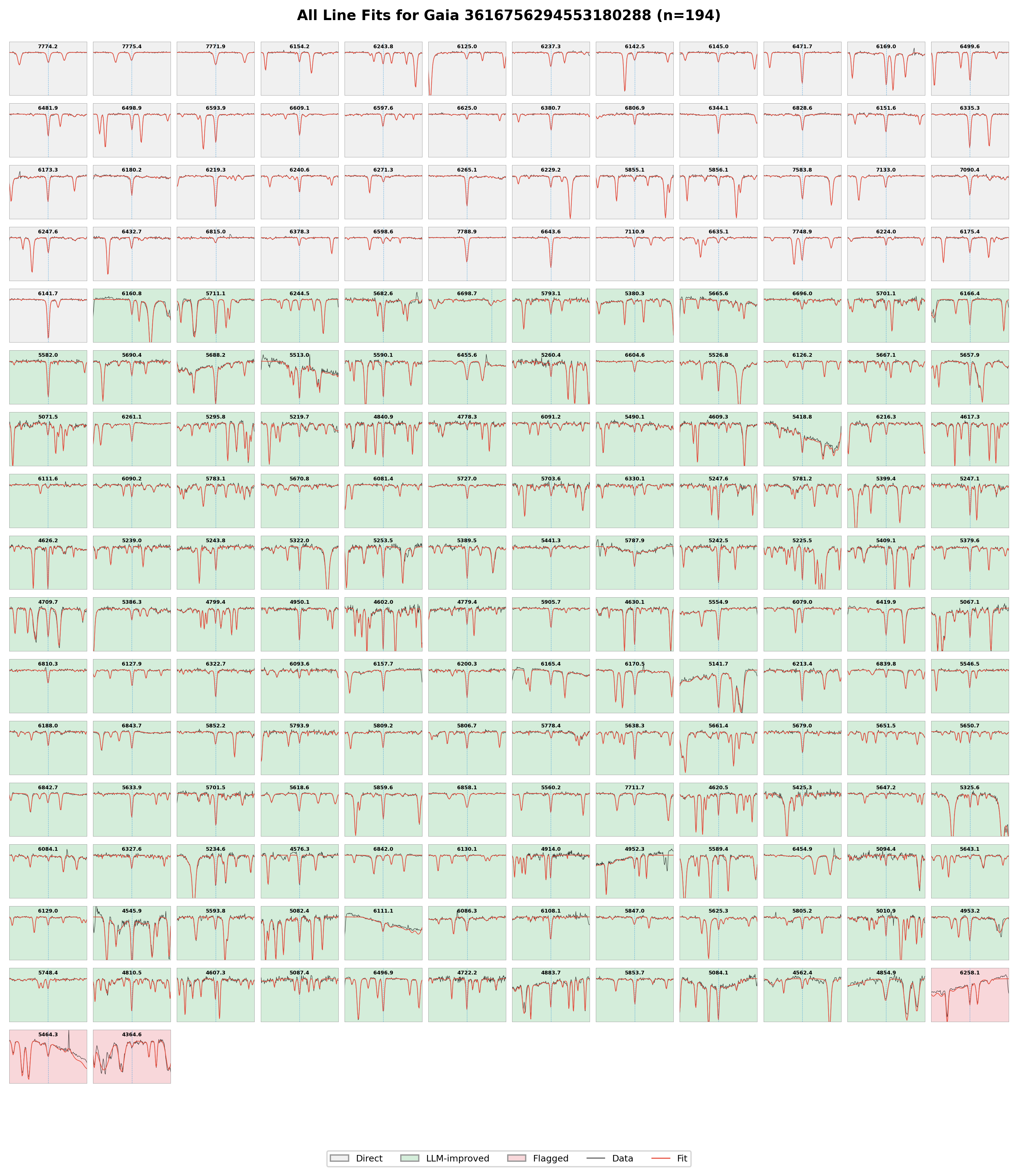}
\caption{All line fits for a second representative spectrum. Background colors: gray = direct fit accepted, green = LLM-improved, red = flagged. Lines are sorted by fit type.}
\label{fig:allfits2}
\end{figure*}

\begin{figure*}[p]
\centering
\includegraphics[width=\textwidth,height=0.92\textheight,keepaspectratio]{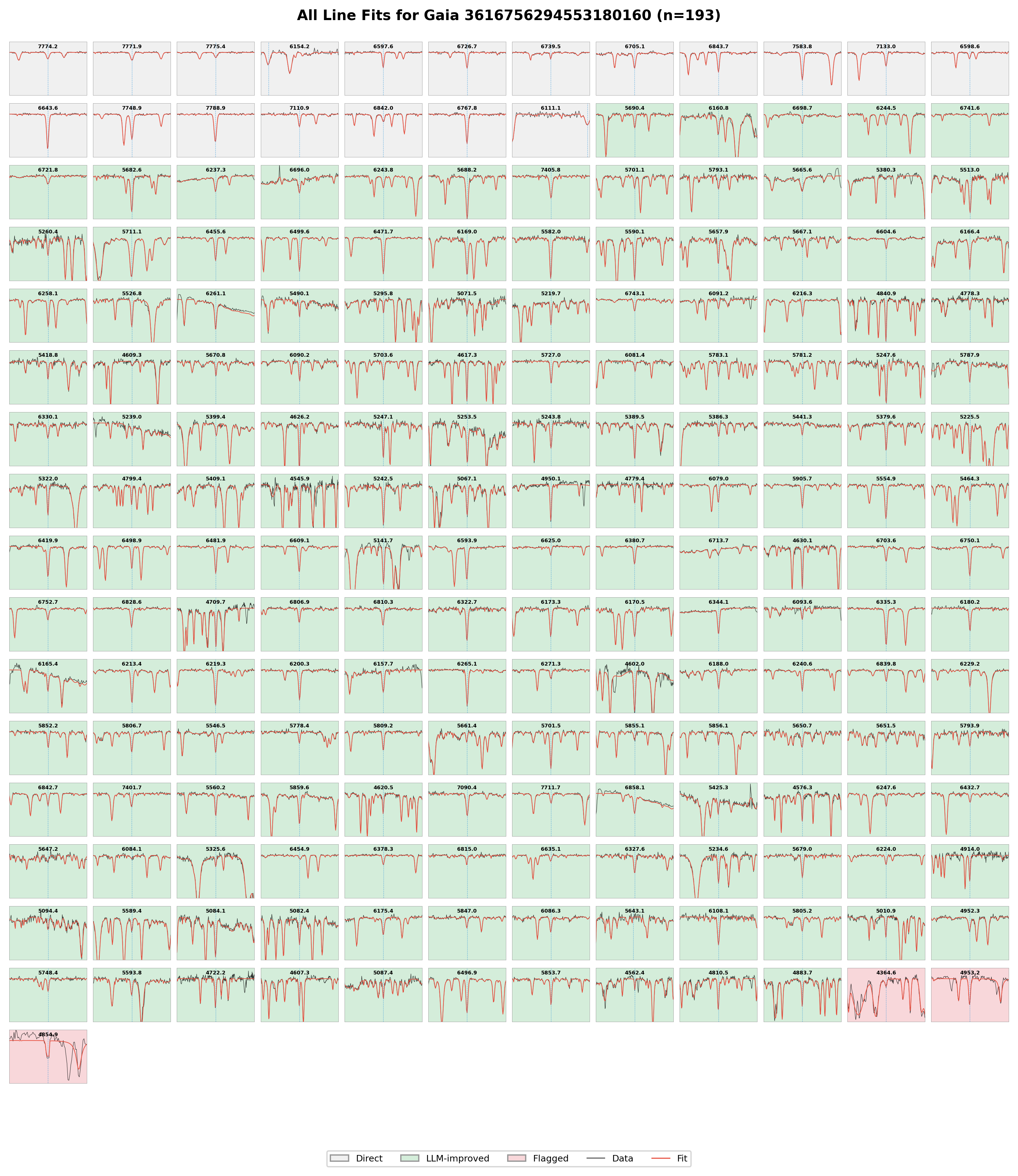}
\caption{All line fits for a third representative spectrum. Background colors: gray = direct fit accepted, green = LLM-improved, red = flagged. Lines are sorted by fit type.}
\label{fig:allfits3}
\end{figure*}

Figures~\ref{fig:allfits1}--\ref{fig:allfits3} show all individual line fits for three representative spectra from the GPT-5-mini sample. Each small panel displays the normalized flux (black) and multi-Voigt model (red), with the target wavelength indicated by the blue vertical line. Background colors distinguish three fit categories: gray indicates direct fits accepted without LLM intervention, green indicates LLM-improved fits, and red indicates flagged lines.

These figures demonstrate that Egent produces high-quality fits across the full wavelength range and line strength distribution. The majority of panels show excellent agreement between data and model, with the fitted Voigt profiles accurately capturing both strong and weak absorption features. Direct fits (gray) dominate in clean spectral regions, while LLM intervention (green) was required for more complex cases involving blends, continuum curvature, or edge effects. The flagged lines (red) represent cases where no satisfactory fit could be achieved during the iterative process---typically due to severe blending, contamination, or low SNR.

Rarely, problematic fits may evade flagging---visual inspection of these figures reveals occasional cases where the model does not perfectly match the data but was not flagged. The flagging threshold can be made more stringent via prompt adjustment, but this involves a trade-off: overly strict criteria flag many acceptable fits, reducing the final sample size without meaningful improvement in quality. The current calibration (flagging $\sim$10--20\% of lines) represents a reasonable balance, but users with more conservative requirements can adjust the flagging thresholds or perform additional manual review of the retained lines.

A key feature visible in these plots: the normalized flux (raw flux divided by locally-estimated continuum) consistently lies near unity across all panels. This demonstrates the quality of Egent's approach---fitting directly from raw flux spectra with simultaneous local continuum estimation, completely bypassing the need for a separate global continuum normalization step. In contrast, C3PO relies on pre-normalized spectra where continuum placement is performed as a separate, subjective step. The systematic offsets in Figure~\ref{fig:onetoone} therefore reflect C3PO's global continuum choices, not Egent's fitting quality. Egent's single-step approach mitigates this source of systematic error.

\bibliographystyle{aasjournal}
\bibliography{references}

\end{CJK*}
\end{document}